\documentclass[structabstract]{aa}
\usepackage{natbib}
\bibpunct{(}{)}{;}{a}{}{,}

\usepackage{graphicx}
\usepackage{lscape}
\usepackage{aalongtable}
\usepackage{subfigure}
\usepackage{txfonts}
\usepackage{comment}
\usepackage{color}
\usepackage{ifthen}

\newcommand{\Ha}{\mbox{H$\alpha$}}
\newcommand{\Hb}{\mbox{H$\beta$}}

\specialcomment{nccomments}{\begingroup\sffamily\color{blue}}{\endgroup}

\begin{document}

\title{Probing star formation and feedback  in dwarf galaxies
\thanks{Based on observations made with ESO Telescopes at Paranal Observatory under programme ID 079.B-0445.}}

\subtitle{Integral field view of the blue compact galaxy Tololo~1937-423}

\author{L. M. Cair\'os
    \inst{1}
    \and
    J.N. Gonz\'alez-P\'erez
    \inst{2}
  }
\institute{Institut f{\"u}r Astrophysik, Georg-August-Universit{\"a}t,
Friedrich-Hund-Platz 1, D-37077 G{\"o}ttingen, Germany \\
           \email{luzma@astro.physik.uni-goettingen.de}
           \and Hamburger Sternwarte,
            Gojenbergsweg 112,
21029 Hamburg, Germany
\email{jgonzalezperez@hs.uni-hamburg.de}
}

\date{Received ..., 2013; accepted ..., 2013}

\abstract 
{Blue compact galaxies (BCG) are gas-rich, low-mass, small systems that
form stars at unusually high rates. This makes them excellent laboratories for investigating the process of star-formation
(SF)
 at 
galactic scales and the effects
of massive stellar feedback on the 
interstellar (and intergalactic) medium.}
{We analyzed the BCG
Tololo~1937-423 using optical integral field spectroscopy to   
probe
 its morphology, stellar content, nebular excitation and ionization properties, and  
 the kinematics of its warm ionized gas.}
{Tololo~1937-423 was observed with the Visible
Multi-Object Spectrograph at the Very Large Telescope. We took data in the wavelength range 
4150-7400\,\AA, covering a field of view of
27$\arcsec \times$ 27$\arcsec$ on the sky with a spatial sampling of 
$0\farcs$67. From these data we built maps in the continuum and 
brighter emission lines, diagnostic line ratio maps, and velocity dispersion fields.  We also generated  the integrated spectrum of
 the main
\ion{H}{ii} regions and young stellar clusters to determine reliable physical
parameters and oxygen abundances.}
{We found that Tololo~1937-423 is currently 
undergoing an extended starburst.
In the H$\alpha$ maps we identified nine major clumps, 
aligned mostly northeast-southwest, and 
stretching to galactocentric
distances $\geq$~2~kpc.  The galaxy presents a single continuum peak that is not cospatial with any knot in
emission lines, indicating at least two relatively recent
 episodes of SF. 
The inhomogeneous dust distribution reachs its maximum [E(B-V)$\sim$0.97]  roughly at the position 
of the continuum peak. We found shocked regions in the galaxy outer regions and at the edges 
of the SF knots. 
The oxygen abundance, 12+log(O/H)$\sim$8.20$\pm$0.1, is similar in all the SF regions, 
suggesting a chemically homogeneous ionized interstellar medium over spatial scales of several kpc.
The ionized gas kinematics displays an overall regular rotation around a 
northwest-southeast axis, with a maximum velocity of 70$\pm$7~km~s$^{-1}$.}
{The morphology of the galaxy and the two different episodes of SF
suggest a scenario of triggered (induced by supernova shock waves) SF acting in Tololo~1937-423. 
The inferred 
ages for the
different SF episodes ($\sim$13-80~Myr for the central post-starburst and 5-7~Myr for the
ongoing SF) are  consistent with triggered SF, with the most recent SF episode caused by the
collective effect of stellar winds and supernova explosions from the central post-starburst. The velocity
dispersion pattern, with higher velocity dispersions found at the edges of the SF regions, and shocked regions in the galaxy, also favor this scenario.}

\maketitle
%

\section{Introduction}

Blue compact galaxies (BCG) have gained ample popularity in extragalactic research, and 
for
good reason, because the study of these small, gas-rich and mostly metal-poor objects
provides
insights into fundamental questions in contemporary extragalactic astronomy. Because of their low
luminosity, compactness (M$_{B}\geq$-18; starburst diameter $\leq$~1~kpc;
\citealp{ThuanMartin1981,Cairos2001b}), and their unusually high star-forming rates (up to
3~M$\odot$~yr$^{-1}$,  \citealp{Fanelli1988,HunterElmegreen2004}), BCGs are excellent
laboratories to investigate the star formation (SF) process in galaxies and the effect of stellar feedback
on the surrounding interstellar medium (ISM).

Blue compact galaxies also lack the complex gas dynamics of spirals, that is, neither density
waves nor shear forces are at play, which means that the SF process can be studied in a
relatively simple environment \citep{Hunter1997,HunterElmegreen2004}.  In addition, 
these low-mass galaxies have a shallower gravitational potential and lower escape velocities
than typical spirals, therefore the effects of feedback processes can be very dramatic. It
has indeed been suggested that BCGs could eject a significant fraction of the enriched matter
that is returned by supernovae (SN) and winds into the intergalactic medium (IGM;
\citealp{Dekel1986,MacLow1999}), and this feedback has been employed to explain some questions that are still
open to discussion, such as the absence of very low-mass galaxies after reionization
\citep{Barkana1999}, the slope of the galaxy luminosity function \citep{Dekel2003}, or the
mass metallicity relation \citep{Larson1974}.


The importance of these questions prompted us to initiate an integral field spectroscopy (IFS)
analysis of a BCG sample, aiming primarily at understanding their recent SF and the
effect of the massive stellar population on the surrounding ISM 
\citep{Cairos2009a,Cairos2009b,Cairos2010,Cairos2012,Cairos2015}. The first data products from
this work, which included emission and diagnostic line maps, interstellar extinction maps and
velocity fields, generated interesting results and some intriguing new questions, which necessitated further work. As a consequence, our next step has been to select a
subsample of objects on the basis of our results, which is to be the subject of a series of
individual studies. Investigations of large samples are essential to open a new research
field, establish general trends and find correlations among fundamental parameters, but
detailed analyses of individual objects are needed to better understand the physical
processes underlying galaxy  parameters and behaviors. 

This is the second in the series of papers presenting the analysis of individual galaxies; results for the BCG Haro14 have
been published recently \citep{Cairos2017}. Here we investigate the BCG Tololo~1937-423, an object that has not  received
much attention so far. It is included in the spectrophometric catalogs of \ion{H}{ii} galaxies by \cite{Terlevich1991} and
\cite{Kehrig2004} and in the study of stellar populations in a sample of \ion{H}{ii} galaxies by \cite{Westera2004}.  Surface
photometry in B and R   has been presented in \cite{Doublier1999,GildePaz2003}, and \cite{GildePaz2005}. Broadband frames of
the galaxy reveal a central high-surface brightness (HSB) region with blue colors ($B-R \approx$ 0.6, \citealp{Doublier1999}),
from which conspicuous extensions depart toward the northeast and south, in addition to a more extended, red, and regular low
surface brightness (LSB) component (see Figure~\ref{Figure:Tololo1937-FOV}). H$\beta$ narrowband imaging, unfortunately of a
limited quality, has been published by \cite{Lagos2007}.

Tololo~1937-423 was part of our IFS study of a BCG sample \citep{Cairos2015}, which revealed it as an
intriguing object. In particular,  the size and morphology of its current episode of SF singled
out this galaxy as a perfect target for a further, deeper analysis. Tololo~1937-423 presents a
spatially extended episode of SF that is taking place in a number of knots distributed mainly 
northeast-southwest, in a structure larger than  4~kpc. Curvilinear features resembling bubbles depart
from the galaxy center to the northeast and southwest, and smaller SF knots, located in these
filaments, suggest an episode of triggered SF. The curvilinear structures and filaments,
and the high values of [\ion{S}{ii}]~$\lambda\lambda6717,\,6731$/\Ha\ in the galaxy outskirts, which are
indicative of shocks, reveal a perturbed ISM. The origin of the
perturbation is most probably the collective action of stellar
winds and SN explosions.

\begin{table}
\caption{Basic parameters for Tololo~1937-423.}
\begin{center}
\begin{tabular}{lcc}
\hline\hline
Parameter   & Data & Reference \\ 
\hline
 RA (J2000)                        &  19$^h$40$^m$58$\fs$6  &      \\
 DEC (J2000)                       & -42$\degr$15$\arcmin$45$\arcsec$     &      \\
 Helio. radial velocity            &  2754$\pm$12 kms$^{-1}$ &     \\
 Distance                          &  41.1$\pm$2.9      Mpc  &      \\
 Scale                             & 199~pc$\arcsec$$^{-1}$  &     \\
 A$_{B}$          & 0.284                           &        \\
 R$_{25,B}$                         & 2.4 kpc                   & D99\\
 m$_{B}$          & 15.14$\pm$0.04$^{a}$ mag          &  GP03 \\
 m$_{R}$          & 14.25$\pm$0.03$^{a}$ mag          &  GP03 \\
 M$_{B}$          & -17.92$\pm$0.03$^{b}$ mag         &  \\
 \hline
\end{tabular}
\end{center}
Notes: RA, DEC, heliocentric velocity, distance, scale, and Galactic extinction 
taken from NED 
(http://ned.www.ipac.caltech.edu/). The 
distance is calculated using a Hubble constant of 73 km s$^{-1}$ Mpc$^{-1}$,
and taking into account the influence of the Virgo cluster, the Great
Attractor, and the Shapley supercluster. 
(a) Integrated magnitudes from
\citep{GildePaz2003}, corrected for Galactic extinction; (b) absolute magnitud in the B-band
computed from the tabulated B-magnitude and distance.
References: D99 = \cite{Doublier1999}; GP03 = \cite{GildePaz2003}.
\end{table}

\section{Observations and data processing}

\label{thedata}

\subsection{Data collection and reduction}

Tololo~1937-423 was observed at the Very Large Telescope (VLT; ESO Paranal Observatory, Chile)  with the
\emph{Visible Multi-Object Spectrograph} (VIMOS; \citealp{LeFevre2003}) in its integral field unit (IFU)
mode. The observations were made in Visitor Mode in August 2007 with the blue (HR-blue) and orange
(HR-orange) grisms in high-resolution mode (dispersion of 0.51\,\AA\,pix$^{-1}$ in the wavelengths range
of 4150--6200\,\AA, and dispersion of 0.60\,\AA\,pix$^{-1}$  in the range 5250--7400\,\AA). A field of
view (FoV) of 27$\arcsec \times$ 27$\arcsec$ on the sky was mapped with a spatial sampling of 
$0\farcs$67.

The integration time was 4320~s both with the HR-blue and with the HR-orange grism. The weather
conditions were good: the seeing ranged from 0.51-0.93~arcsec when we observed the red part of the
spectrum, and from 0.67-1.58~arcsec in the blue. The spectrophotometric standard EG~274 was
observed for flux calibration. 

The data were processed using the ESO VIMOS pipeline (version 2.1.11) via the graphical user
interface {\sc Gasgano}. The observations and a complete description of the data
reduction have been presented in \cite{Cairos2015}.

\subsection{Building the galaxy maps}
\label{Section:mapas}

We computed the fluxes of the brightest emission lines in each individual spectrum by fitting Gaussian
functions to the line profiles.  The lines were fit with the {\tt Trust-region}
algorithm for nonlinear least squares, using the task {\tt fit} of {\sc Matlab}. The lines H$\beta$,
[\ion{O}{iii}]$\lambda\lambda~4959,5007$, H$\alpha$, [\ion{N}{ii}]$\lambda\lambda~6548,6584,$ and 
[\ion{S}{ii}]$\lambda\lambda~6717,6731$ were measured in the spectrum of each spatial element (spaxel).
Along with the line flux, the fitting provides the centroid position, line width, continuum,
and their corresponding uncertainties. 

A single-Gaussian profile fit was used in all but in the  H$\beta$ line.  Direct measurements of the
emission fluxes of the Balmer lines underestimate the real value, since their fluxes are
affected by underlying stellar absorption \citep{McCall1985,Olofsson1995}.  The
underlying stellar component, often visible in the spectra as pronounced absorption wings around the
lines in emission, adds to the emission flux, decreasing its final value; this absorption can be
significant: single stellar populations models predict equivalent width values of between 2 and 15 \AA\ 
\citep{Olofsson1995,GonzalezDelgado1999b}. To  account for this effect in H$\beta$, where absorption
wings are clearly seen, we fit the line profile using two Gaussian functions (one in emission and one in
absorption) and derived the absorption and emission fluxes simultaneously. In H$\alpha$ the absence of
absorption features in the observed spectrum makes a reliable decomposition impossible, and an alternative
approach must be adopted. We assumed the equivalent width in absorption in H$\alpha$ to be equal to the absorption in 
H$\beta$, an assumption well supported by the predictions of evolutionary synthesis models
\citep{Olofsson1995}.

A spatial smoothing was applied to increase the accuracy of the fits in regions of low surface
brightness. Depending on the signal-to-noise ratio (S/N) of the spaxel, the closest 5, 9, or 13 spaxels
were averaged before the fit was carried out. This procedure preserves the spatial resolution of the
bright regions of the galaxy while providing a reasonable S/N in the faint parts, although at a lower
spatial resolution.

The parameters derived from the fit (line fluxes, centroid position, line width, and continuum)  were then used to
build the bidimensional maps of the galaxy, taking advantage of the fact that the combined VIMOS data are arranged in a
regular 44x44 matrix.  

The  continuum map was produced by summing over the whole spectral range, but masking the spectral
regions with significant contributions of emission and telluric lines.

Line ratio maps for lines falling within the wavelength range of either grism were derived by simply
dividing the corresponding flux maps. The H$\alpha $/H$\beta$ map was derived after registering and
shifting the H$\alpha$ map to spatially match the H$\beta$ map. The shift was calculated aligning the
center of the brighter \ion{H}{ii}-regions. 

We considered only spaxels with a flux level higher than 3$\sigma$  when we built the final maps.
The H$\alpha$/H$\beta$ map (see Section~\ref{Section:extinction}) was employed to deredden all the
emission line flux maps on a spaxel to spaxel basis.

\begin{figure}
\centering
\includegraphics[angle=0, width=0.8\linewidth]{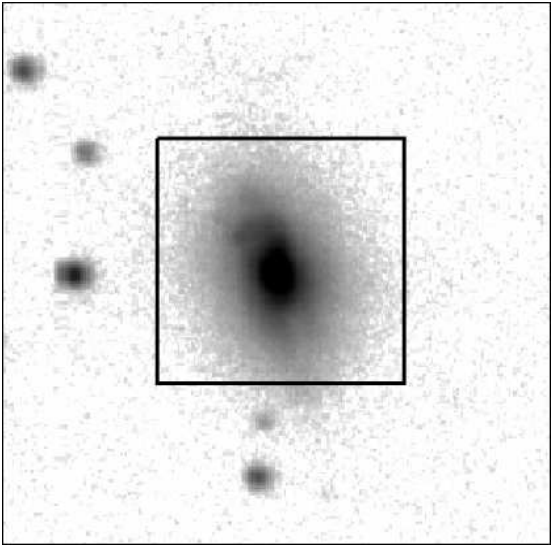}
\caption{B-band image of Tololo~1937-423. The FoV is 1 arcmin; the 
$27\arcsec \times 27\arcsec$ VIMOS FoV is overplotted.
North is up and east to the left. The image, retrieved from 
NED, has been published by \cite{GildePaz2003}}
\label{Figure:Tololo1937-FOV} 
\end{figure}

\section{Results}
\label{Section:analysis}

Figure~\ref{Figure:Tololo1937-FOV} shows the B-band image of Tololo1937-423, with the  $27\arcsec \times
27\arcsec$ VIMOS FoV overplotted. At the galaxy distance, 41.1~Mpc, the spatial scale is
199~pc~arcsec$^{-1}$, with VIMOS we also map about 5.4$\times$5.4~kpc$^{2}$, with a spatial
resolution of about 133~pc per spaxel.

\subsection{Stellar and warm ionized gas distribution}
\label{Section:starsandgas}

\begin{figure}
\centering
\includegraphics[angle=0, width=0.9\linewidth]{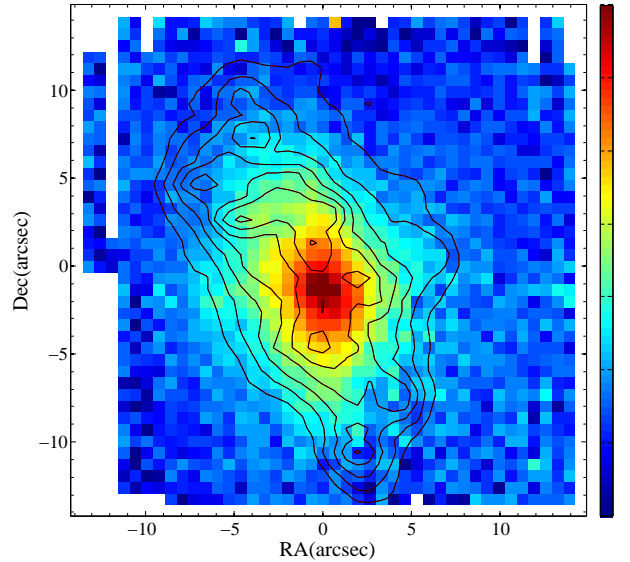}
\caption{Tololo~1937-423 continuum map constructed by summing over the whole orange spectral range, 
but masking the emission lines. Contours in H$\alpha$ flux are overplotted.
The image is scaled in arbitrary flux units.
The FoV is $\sim$ 5.4$\times$5.4~kpc$^{2}$, with a spatial
resolution of about 133~pc per spaxel; north is up and east to the left, 
also in all the maps shown from here on.}
\label{Figure:Tololo1937-co} 
\end{figure}

\begin{figure}
\centering
\includegraphics[angle=0, width=0.9\linewidth]{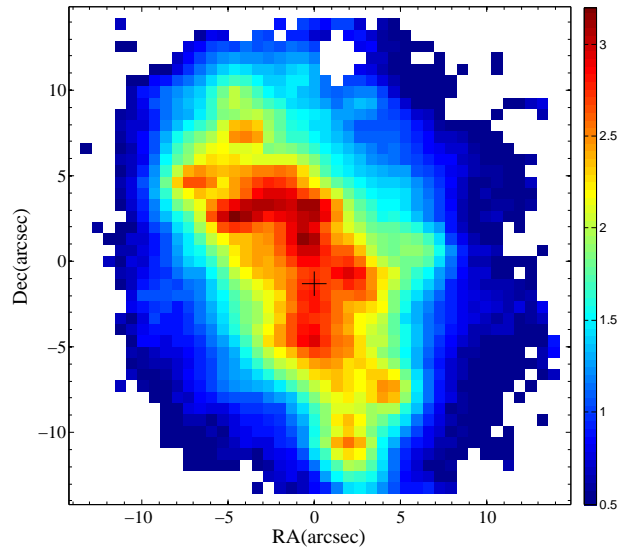}
\caption{H$\alpha$ emission line flux 
map for Tololo~1937-423 (flux units are $10^{-18}$\,erg\,s$^{-1}$\,cm$^{-2}$). The cross indicates
  the position of the continuum peak.}
\label{Figure:Tololo1937-ha} 
\end{figure}

\begin{figure}
\centering
\includegraphics[angle=0, width=0.9\linewidth]{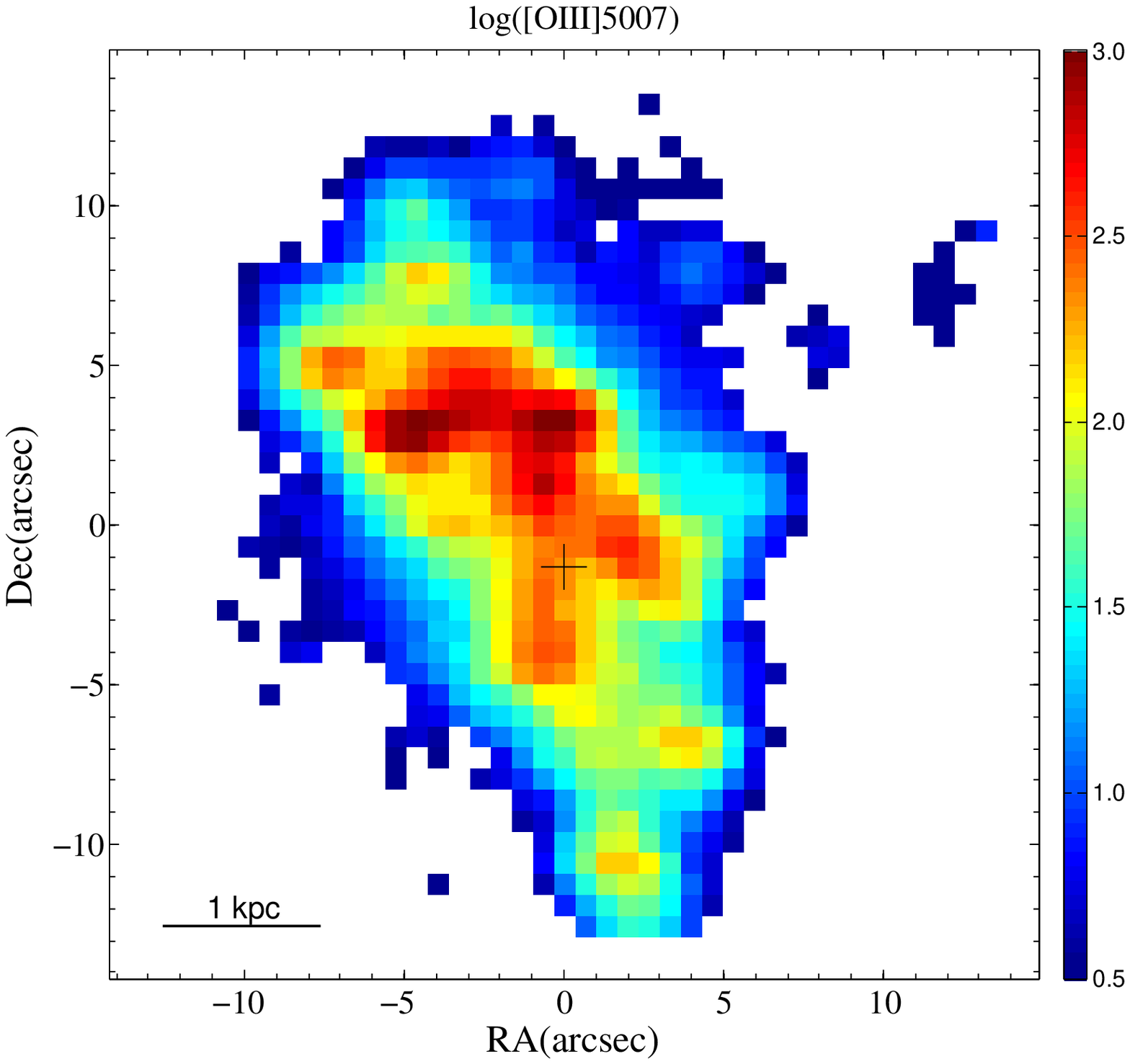}
\caption{[\ion{O}{iii}]$\lambda$5007 emission line flux 
map for Tololo~1937-423. 
}
\label{Figure:Tololo1937-oiii} 
\end{figure}

\begin{figure}
\centering
\includegraphics[angle=0, width=0.9\linewidth]{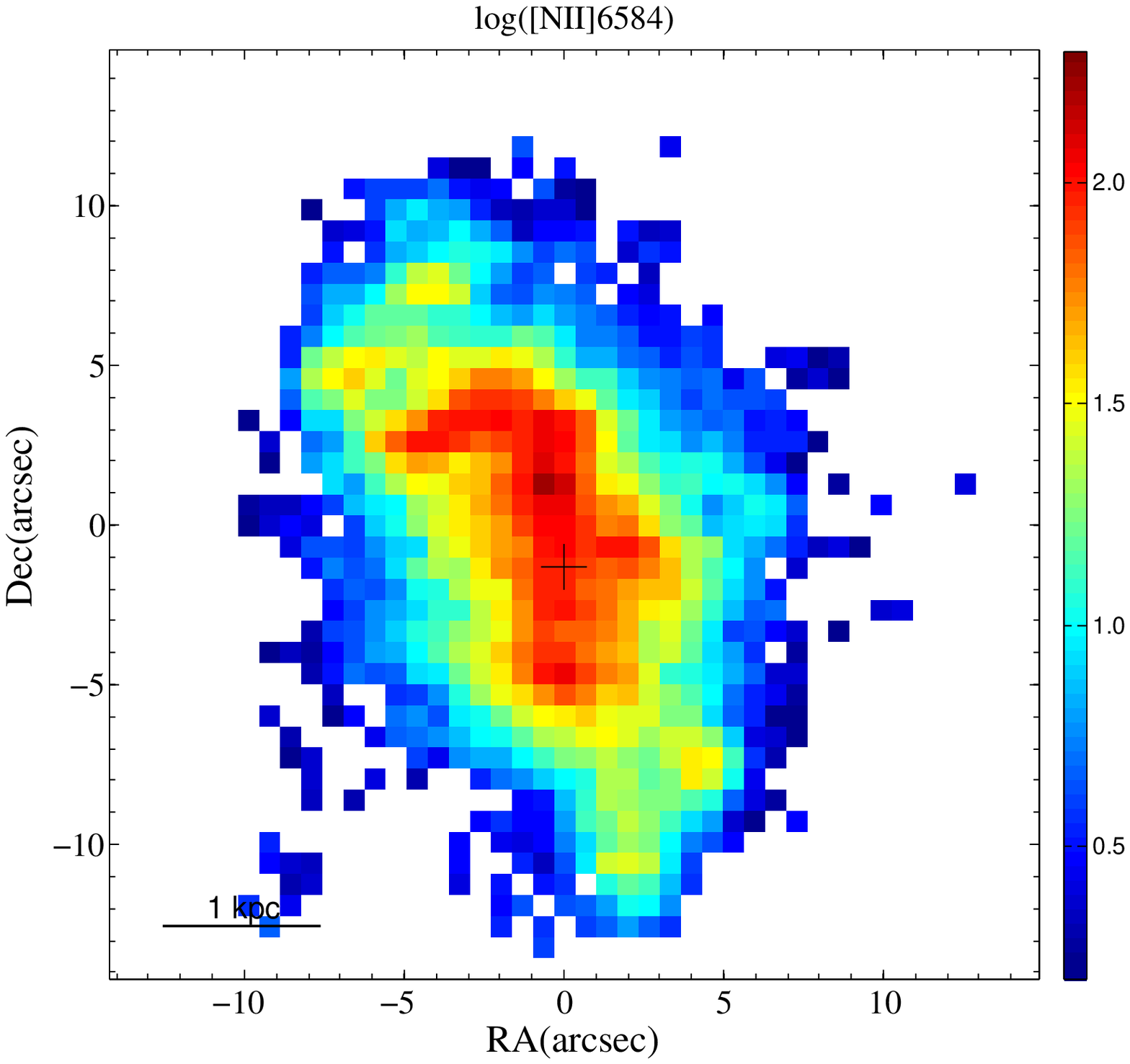}
\caption{[\ion{N}{ii}]$\lambda$6584 emission line flux 
map for Tololo~1937-423. 
}
\label{Figure:Tololo1937-nii} 
\end{figure}

\begin{figure}
\centering
\includegraphics[angle=0, width=0.9\linewidth]{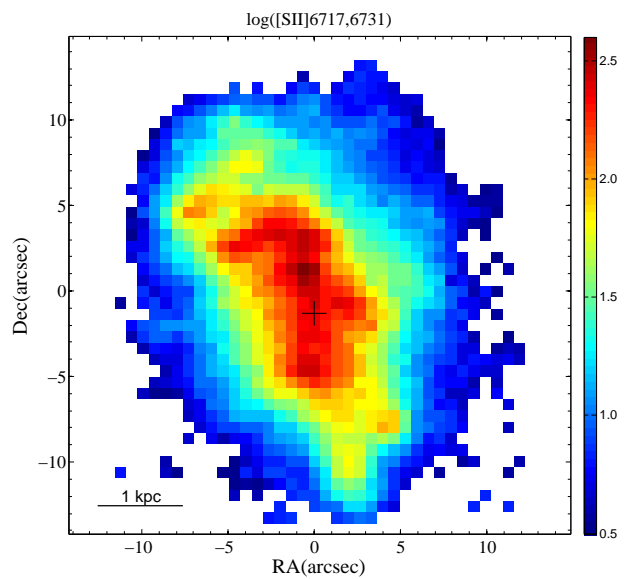}
\caption{[\ion{S}{ii}]$\lambda\lambda$6717,6731 emission line flux 
map for Tololo~1937-423. 
}
\label{Figure:Tololo1937-sii} 
\end{figure}

The continuum intensity of Tololo~1937-423 increases steeply toward the inner part of the galaxy,
where it peaks (Figure~\ref{Figure:Tololo1937-co}). The morphology is distorted in the whole surface
brightness range, most probably because the
extended  SF episode is superposed, and the intensity maximum is situated roughly at the center of the outer
isocontours.  

\begin{figure}
\centering
\includegraphics[angle=0, width=0.8\linewidth]{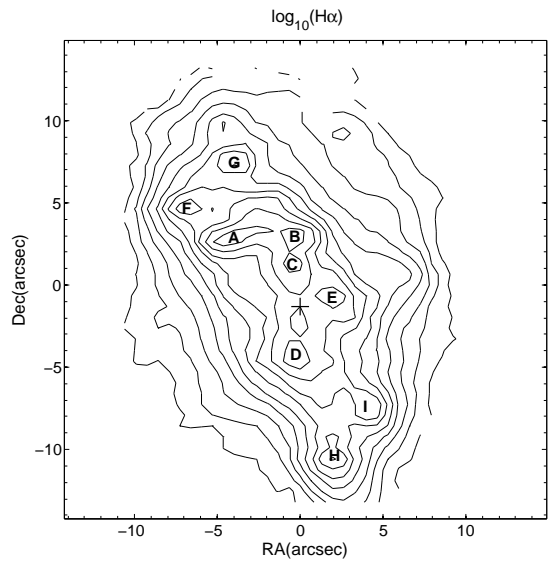}
\caption{Contour map of the H$\alpha$ emission in Tololo~1937-423, with the main SF 
regions labeled.}
\label{Figure:Ha-contour} 
\end{figure}

The distribution of the warm ionized gas, as revealed by the emission line maps, is more knotty
(see~Figure~\ref{Figure:Tololo1937-ha}). An ensemble of SF regions is distributed along an axis  $\sim
30\degr$~NE---the main SF regions are labeled in Figure~\ref{Figure:Ha-contour}. Three bright knots,
namely knots~{\sc a}, {\sc b,} and {\sc c}, are located north of the continuum peak, forming a
curvilinear (arc) structure with a radius of about 530~pc.  Knot~{\sc a}, the peak of the Balmer
line emission, is displaced by about 6.2~$\arcsec$ (1.2~kpc) to the northeast of the continuum peak. From the
central SF regions, filaments depart southwest and northeast, at about 2.0 and 2.5~kpc, respectively. 
Four knots ({\sc d}, {\sc e}, {\sc h,} and {\sc i}) are well visible along the filaments stretching to the southwest, and two other knots ({\sc f} and {\sc g}) are detected in the filaments extending northeast of the peak. Indications of faint filaments are visible to the west. Diffuse H$\alpha$ emission fills almost
the entire FoV.

All the emission line maps display a roughly similar pattern (a complete set of maps for
Tololo~1937-423 can be found in \citealp{Cairos2015}), but the intensity maximum depends on the
wavelength range: in the recombination lines H$\alpha$ and H$\beta$, the emission peaks in knot~{\sc a},
while the high-ionization forbidden line [\ion{O}{iii}] peaks in knot~{\sc b}, and the
low-ionization [\ion{N}{ii}] and [\ion{S}{ii}] lines peak in knot~{\sc c}
(see Figures~\ref{Figure:Tololo1937-ha} to \ref{Figure:Tololo1937-sii}).

Tololo~1937-423 displays a different morphology in the emission line and continuum maps. While the
continuum emission is centrally concentrated and presents a single maximum, emission line maps are more
extended, and nine main SF regions have been identified, none  of them at the spatial position of
the continuum  peak.  Because hydrogen recombination-lines and forbidden lines of metals are generated
in the ionized interstellar gas, and because the continuum emission is mostly produced in the photosphere of
the stars, the comparison of the emission lines and continuum pattern  provides a first glance at the
stellar populations. We find at least two  different stellar components in Tololo~1937-423, a spatially
extended,  very young population that ionizes the surrounding gas and is therefore younger than 
$\sim$10~Myr, and a more evolved and centrally concentrated population.

This morphology suggests that a mechanism of induced SF acts in the galaxy, where an older episode of SF,
which took place in the HSB central regions,  could have triggered the younger SF episode detected in
emission lines.   That knots are distributed in curvilinear structures
and the distorted
emission line morphology also indicate an ISM that is perturbed by the actions of SN
explosions.  In Section~\ref{discuss}  we investigate such a scenario.

\subsection{Line-ratio maps}

Specific line ratios can be used to derive information about the dust content, the physical parameters,
and the excitation and ionization mechanisms acting in a nebula \citep{Aller1984,Osterbrock2006}. The
main advantage of integral field data is that they permit computing  these line ratios for every spatial
resolution element, to build maps, and from them,  to derive the dust, physical parameters, and excitation
and ionization  distribution across the observed FoV, and in particular, to investigate and quantify
their possible spatial variations.

\subsubsection{Interstellar extinction pattern}
\label{Section:extinction}

In the optical, the interstellar extinction can be computed from the ratio of the fluxes of the 
Balmer lines (see also Section~\ref{Section:integrated}).  Although H$\gamma$ falls into the observed
spectral range, we used only the H$\alpha$/H$\beta$ ratio to trace the extinction pattern because
H$\gamma$, significantly fainter than H$\alpha$ and H$\beta$, is also more severely affected by the
underlying stellar absorption \citep{Olofsson1995}. 

\begin{figure}
\centering
\includegraphics[angle=0, width=0.9\linewidth]{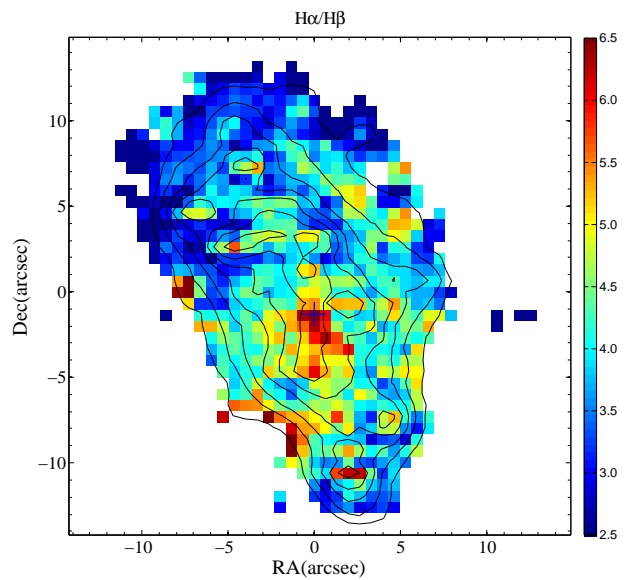}
\caption{H$\alpha$/H$\beta$ ratio map for Tololo~1937-423 with the contours of the H$\alpha$ flux
map overplotted.}
\label{Figure:reddening} 
\end{figure}

The H$\alpha$/H$\beta$ line-ratio map of Tololo~1937-423 is displayed in
Figure~\ref{Figure:reddening}. The galaxy shows an inhomogeneous extinction pattern, with the
maximum situated roughly at the position of the continuum peak, where
H$\alpha$/H$\beta$ reaches values of up to 6.5 (which translate into an extinction coefficient,
C(H$\beta$)=0.96, and a color excess, E(B-V)=0.71). The extinction is relatively high at or very
close to  the position of all SF regions (H$\alpha$/H$\beta$ $\geq$4), while ratios close to the
theoretical value (H$\alpha$/H$\beta$ $\geq$2.7 for  case~B with a T=10000~K and
N$_{e}\leq$100~cm$^{-3}$, \citealp{Osterbrock2006}) are reached at larger galactocentric distances.

The high values of the insterstellar extinction and its significant spatial variability agree with
previous results for BCGs investigated by means of IFS \citep{Cairos2009a, Cairos2009b, Cairos2015,
Lagos2009, Lagos2016, James2013, Cairos2017}, and stress  the importance of information on the dust
spatial distribution, even when dealing with small and metal-poor systems such as BCGs. Assuming a
single spatial constant value for the extinction, as is usually done for long-slit
spectroscopic observations, can lead to significant errors in the fluxes and magnitudes in different
galaxy regions. In the case of Tololo~1937-423,  for instance, if we   were to correct the photometry of
knot~{\sc b} with the extinction coefficient derived for the nuclear region, we would derive a $(B-V)$
color $\sim$0.31~mag bluer and a H$\alpha$ flux 2.3 times higher; for knot~{\sc g}, we would obtain a
$(B-V)$ color $\sim$0.40~mag bluer and a H$\alpha$ flux three times higher. These errors would translate into
large errors in ages and star formation rates (SFRs). Uncertainties $\geq$~0.30~mag in the $(B-V)$
color of a young stellar population make this observable useless as an age indicator, as the $(B-V)$
variations  are on the same order (0.3-0.4~mag) when the age increases for 0 to 100~Myr. An overestimation
of the H$\alpha$ flux by a factor 3 implies the same factor in the derivation of the SFR.

\subsubsection{Mapping the electron density}

\label{Section:density}

\begin{figure}
\centering
\includegraphics[angle=0, width=0.9\linewidth]{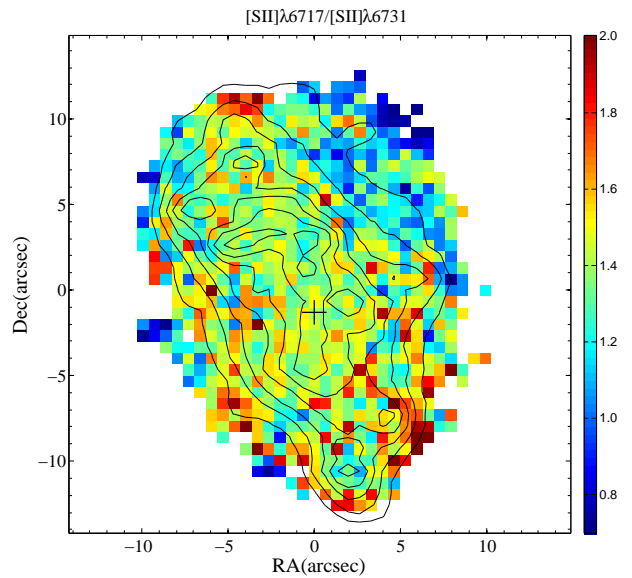}
\caption{Tololo~1937-423 electron density sensitive [\ion{S}{ii}]~$\lambda6717$/[\ion{S}{ii}]~$\lambda6731$ 
emission-line ratio with the H$\alpha$ contour overplotted.}
\label{Figure:density} 
\end{figure}

The ratio of the collisionally excited  [\ion{S}{ii}]~$\lambda\lambda$6717,6731 lines is a sensitive
electron density  ($N_{\rm e}$) diagnostic in the range 100-10000 cm$^{-3}$
\citep{Aller1984,Osterbrock2006}. A map of the ratio of these lines in Tololo~1937-423 is displayed
in  Figure~\ref{Figure:density}.

In the central part of the galaxy, where the SF is taking place, high ratio values
([\ion{S}{ii}]~$\lambda6717$/[\ion{S}{ii}]~$\lambda6731$$\geq$~1.35) are  uniformily found, 
indicative of low electron densities ($N_{\rm e}\leq$100~cm$^{-3}$). In the northwest
regions, where the H$\alpha$ flux is lower, the ratio decreases
([\ion{S}{ii}]~$\lambda6717$/[\ion{S}{ii}]~$\lambda6731\sim$0.9-1.1), indicating  densities
$N_{\rm e}\sim$ 940-630~cm$^{-3}$.  This density pattern of low-density regions in the SF area and an increasing
density toward the galaxy outskirts is consistent with the scenario of a disturbed
ISM, in which higher densities are associated with filaments
that form in the expanding fronts that are
produced by stellar winds and SN explosions. This agrees with the results of 
Section~\ref{Section:diagnostics}: the regions of higher densities coincide with shocked
regions.

\subsubsection{Excitation and ionization structure}

\label{Section:diagnosticmaps}

The two most common power sources responsible for the excitation and ionization of the interstellar
gas in emission-line galaxies are high-energy photons from hot stars or an active galaxy nucleus
(AGN), and shocks caused by SN explosions and massive stellar winds \citep{Dopita2003}. 

Specific emission-line ratios have been proven to be powerful tools to
investigate these excitation mechanisms \citep{Baldwin1981,Veilleux1987}, the
most frequently employed being [\ion{O}{iii}]~$\lambda5007$/\Hb{},
[\ion{N}{ii}]~$\lambda6584$/\Ha, [\ion{S}{ii}]~$\lambda\lambda6717,\,6731$/\Ha,{}
and  [\ion{O}{i}]~$\lambda6300$/\Ha{}.   These ratio maps, except
for [\ion{O}{i}]~$\lambda6300$/\Ha{},  have been constructed for  Tololo~1937-423.
Although  [\ion{O}{i}]~$\lambda6300$ falls within the observed spectral range,
it is severely affected by residuals from the sky subtraction, which hampered a
precise determination of its flux.

\begin{figure}
\centering
\includegraphics[angle=0, width=0.9\linewidth]{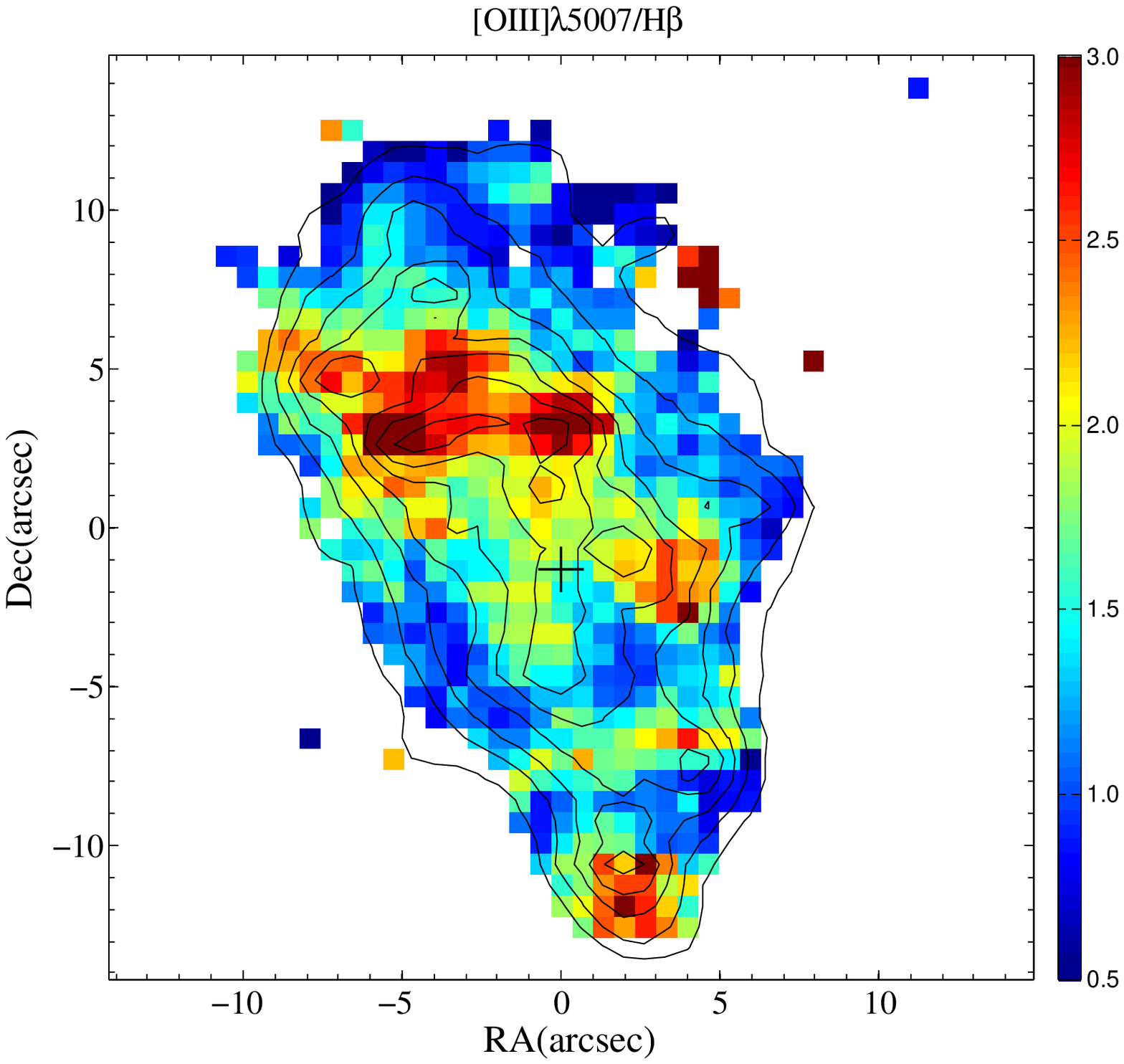}
\caption{[\ion{O}{iii}]~$\lambda5007$/\Hb{} emission-line ratio map for Tololo~1937-423  with contours on 
H$\alpha$ overplotted.}
\label{Figure:diagnostic-oiii} 
\end{figure}

Figure~\ref{Figure:diagnostic-oiii} shows the spatial distribution of the
[\ion{O}{iii}]~$\lambda5007$/H$\beta$ (excitation) map of Tololo~1937-423. High values in this map
indicate  high ionization. As expected, the excitation peaks are associated with SF regions, that is,
with hot (young) massive stars. The excitation values are consistent with 
photoionization by hot stars in the whole FoV, but noticeable variations appear among different SF
knots. The highest excitation ([\ion{O}{iii}]~$\lambda5007$/H$\beta$ up to 3.7) is found in
knots~{\sc a} and~{\sc b}, while in knots~{\sc d}, ~{\sc g} and {\sc i}, the values are considerably
lower (between 1.2 and 1.6). The
[\ion{O}{iii}]~$\lambda5007$/\Hb{} ratio depends mostly on the  hardness of the radiation field (harder
fields produce higher excitation) and on the metallicity (at higher metallicities
the cooling is more effective, the temperature drops, and the excitation diminishes). Since no
significant abundance gradient has been found among the SF knots in Tololo~1937-423 (see
Section~\ref{Section:abundances}), the variations in their excitation is most probably associated with
changes in the radiation field. 

In the northweast of the galaxy and southweast of knot~{\sc e} we can distinguish two regions that do
not coincide with any peak in the SF and where [\ion{O}{iii}]~$\lambda5007$/H$\beta$ also reaches
high values. Such high ionization regions, located in the galaxy outskirts or in the periphery of
SF knots,  are most probably related with the presence of shocks. This is supported by 
Figure~\ref{Figure:map-diagnostic}, which shows that shocks are acting on these areas. 


\begin{figure}
\centering
\includegraphics[angle=0, width=0.9\linewidth]{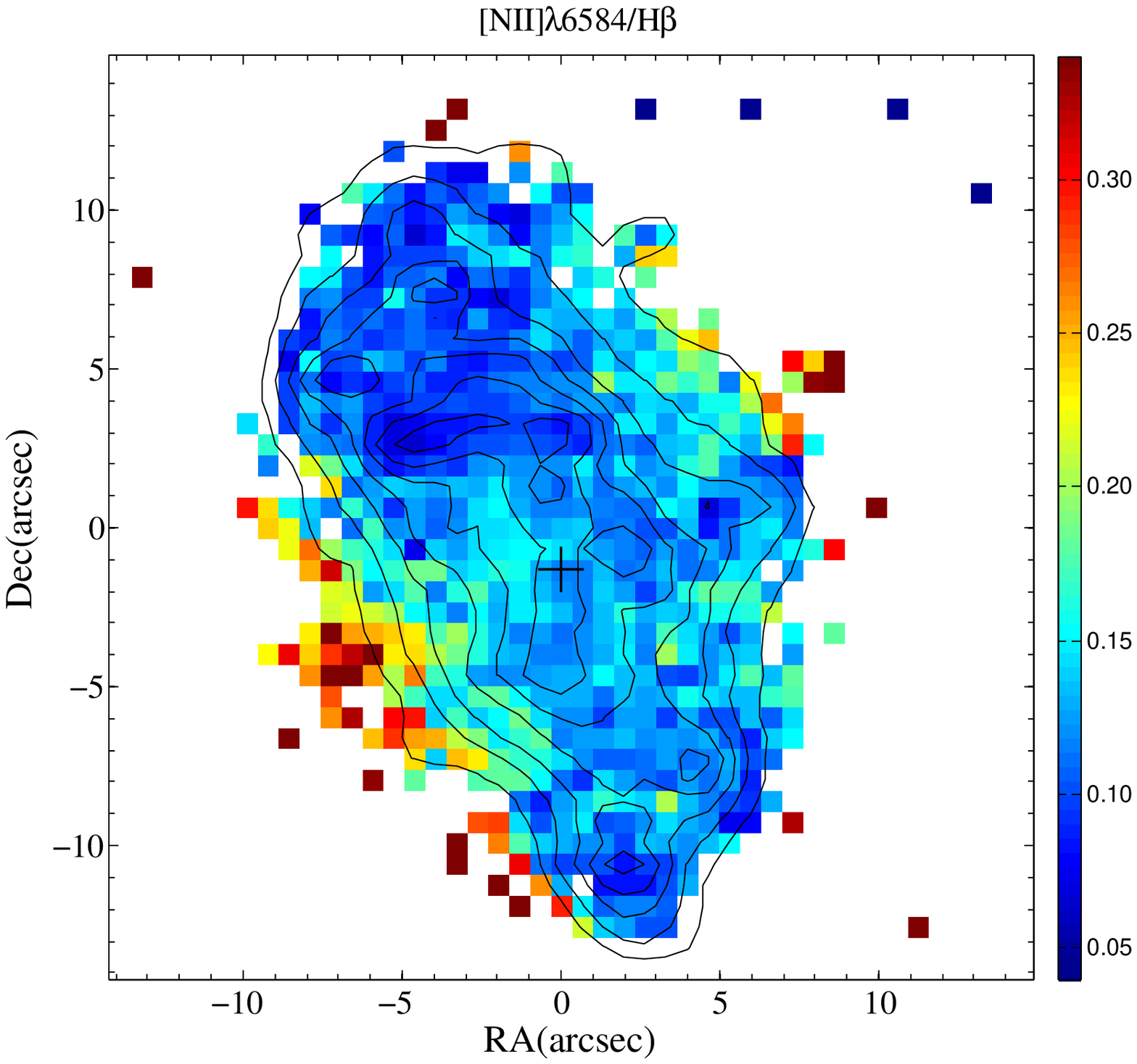}
\caption{[\ion{N}{ii}]~$\lambda6584$/\Ha\ emission-line ratio map for Tololo~1937-423 with contours on 
H$\alpha$ overplotted.}
\label{Figure:diagnostic-nii} 
\end{figure}

\begin{figure}
\centering
\includegraphics[angle=0, width=0.9\linewidth]{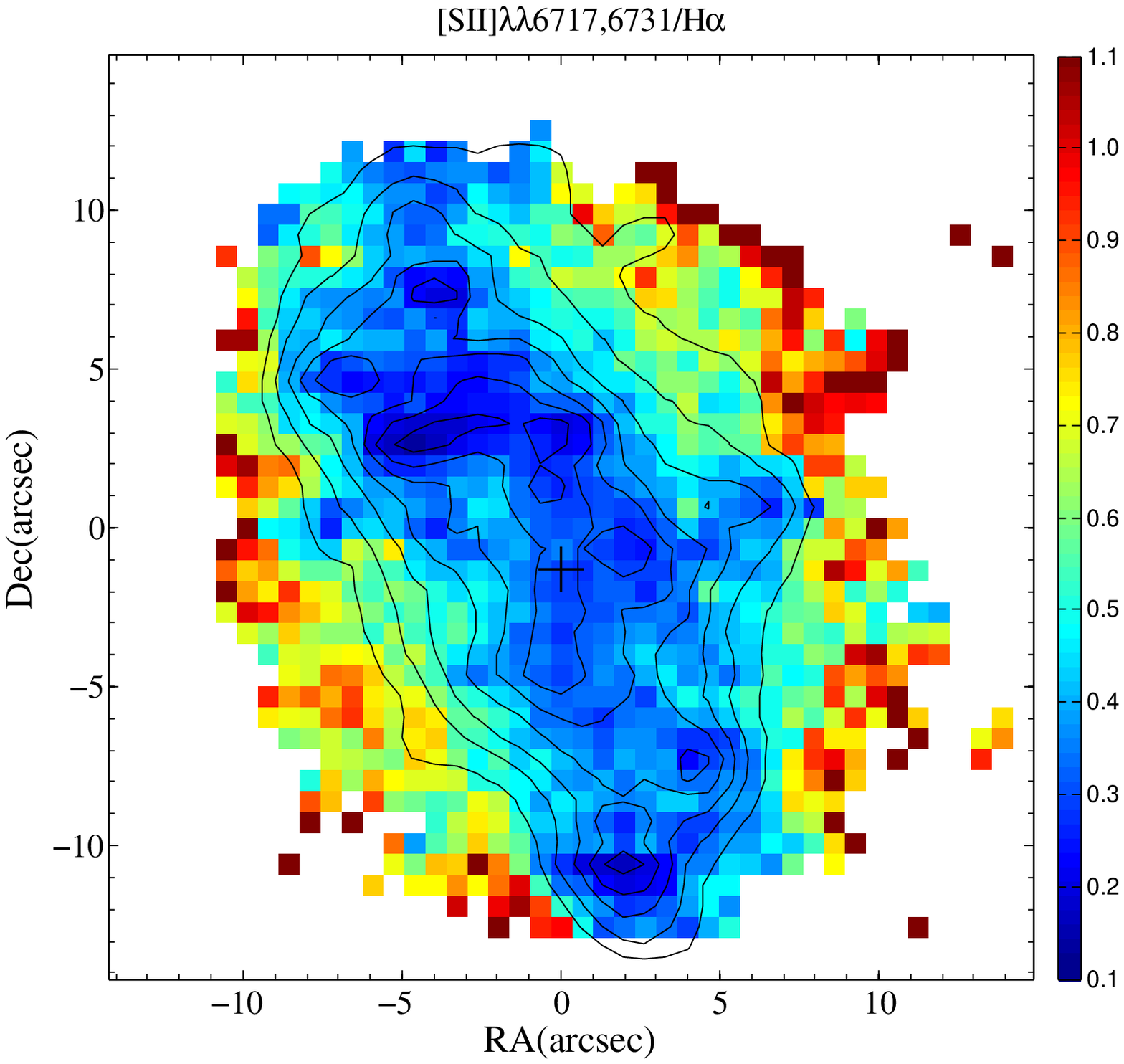}
\caption{[\ion{S}{ii}]~$\lambda\lambda6717,\,6731$/\Ha\ emission-line ratio map for Tololo~1937-423 with contours on 
H$\alpha$ overplotted.}
\label{Figure:diagnostic-sii} 
\end{figure}

The [\ion{N}{ii}]~$\lambda6584$/\Ha\ and [\ion{S}{ii}]~$\lambda\lambda6717,\,6731$/\Ha\  maps
of Tololo~1937-423 (Figures~\ref{Figure:diagnostic-nii} and \ref{Figure:diagnostic-sii}) show basically
the same structure. Like the excitation, both maps trace the SF regions, and the minima in
[\ion{N}{ii}]~$\lambda6584$/\Ha\ and [\ion{S}{ii}]~$\lambda\lambda6717,\,6731$/\Ha\ are associated with
the maxima in the excitation, as expected in regions ionized by UV photons coming from
massive stars. The line-ratio values are consistent with photoionization from hot
stars, except for the outer regions in the [\ion{S}{ii}]~$\lambda\lambda6717,\,6731$/\Ha\ map, where 
large numbers ([\ion{S}{ii}]~$\lambda\lambda6717,\,6731$/\Ha\ $\geq$ 0.8) indicate that another
ionization mechanism (most probably shocks) must be playing a relevant role.

\subsubsection{Diagnostic diagrams}

\label{Section:diagnostics}

Since the pioneering work of \cite{Baldwin1981}, diagnostic-diagrams have been commonly employed to
distinguish among the different excitation mechanisms acting in emission-line galaxies
\citep{Veilleux1987,Kewley2001,Kauffmann2003,Kewley2006}. The advent of integral field spectrographs
working at large telescopes opened new and valuable possibilities in the application of these diagrams.
Integral-field high-sensitivity spectroscopy permits building diagnostic diagrams in terms of spaxels,
that is, to plot in the diagram each individual element of spatial resolution, and thus to
investigate  the existence of distinct mechanims acting in different regions of the galaxy.  This
technique has been  applied in recent years with great success to nearby galaxies
\citep{Sharp2010,Rich2011,Rich2012,Rich2015,Leslie2014,Belfiore2015,Belfiore2016,Cairos2017}.

\begin{figure*}
\centering
\includegraphics[angle=0, width=0.75\linewidth]{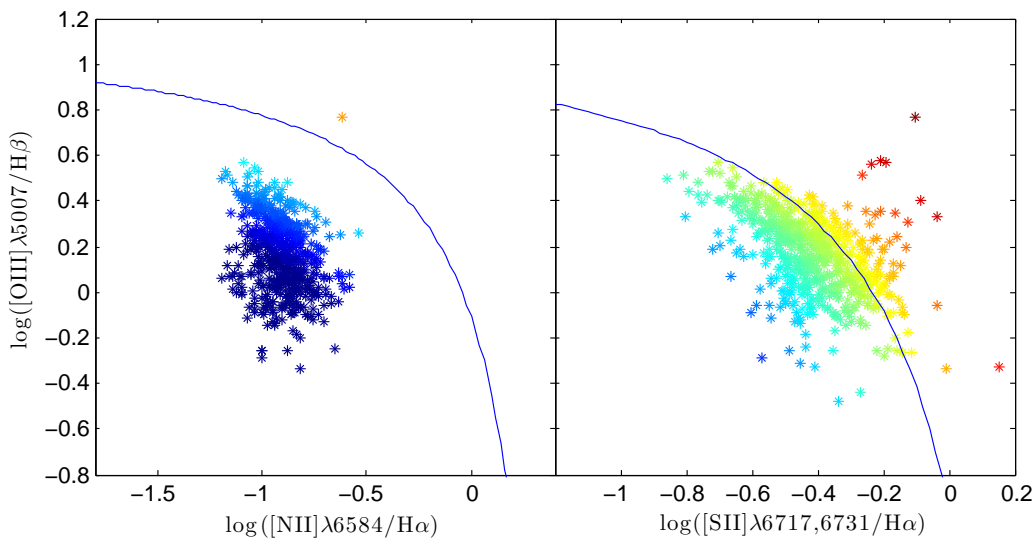}
\caption{Optical emission-line diagnostic
diagram for the different spaxels in  Tololo~1937-423. The theoretical "maximum starburst line"  
derived by \cite{Kewley2001} is also
included in the figure. To better visualize the results in the diagram, the points are
color-coded according to their distance to the maximum starburst line. In both diagrams, the lower left section
of the plot is occupied by spaxels in which the dominant energy source is the
radiation from hot stars (blue points in the figure). Additional ionizing mechanisms
shift the spaxels to the top right and right part of the diagrams (yellow to red colors). 
}
\label{Figure:diagnostic-spaxel} 
\end{figure*}

To investigate the power sources acting in Tololo~1937-423, we plot the values of its  emission-line ratios in the
diagnostic diagrams of [\ion{O}{iii}]~$\lambda5007$/\Hb{} versus  [\ion{N}{ii}]~$\lambda6584$/\Ha\ and
[\ion{S}{ii}]~$\lambda\lambda6717,\,6731$/\Ha\  (see Figure~\ref{Figure:diagnostic-spaxel}). We also
 draw in the graphics the
{\em \textup{maximum starburst line}} or {\em \textup{photoionization line}} derived  by \cite{Kewley2001}. These authors used the starburst
grids from the {\sc pegase}  evolutionary synthesis models \citep{Fioc1997} and the gas ionizing code {\sc mappingsiii}
\citep{Sutherland1993} to parametrize an extreme starburst line, which  marks the limit between gas photoionized by hot stars
and gas ionized via other mechanisms. Accordingly, the flux ratios of any object lying above this boundary cannot be modeled by
hot star photoionization, but require an additional contribution from a harder radiation source such as an AGN or shock
excitation. This maximum starburst line is a conservative border, meaning that it provides a lower limit to the amount of
shock-ionized gas \citep{Kewley2001, Kauffmann2003}.

We found for Tololo~1937-423 that a significant number of spaxels fall out of the areas occupied
by photoionization by stars in the [\ion{O}{iii}]~$\lambda5007$/\Hb{} versus
[\ion{S}{ii}]~$\lambda\lambda6717,\,6731$/\Ha\ diagram.  The ratio
[\ion{N}{ii}]~$\lambda6584$/\Ha{}, weakly dependent of the hardness of the radiation and strongly
dependent on the metallicity, is not effective in separating shocks from photoionized regions. In
particular at low metallicities (0.2Z$\odot\leq$Z$\leq0.4$Z$\odot$), the diagrams are degenerated,
and  shock-ionization and photoionization overlap \citep{Allen2008,Hong2013}.

\begin{figure}
\centering
\includegraphics[angle=0, width=0.8\linewidth]{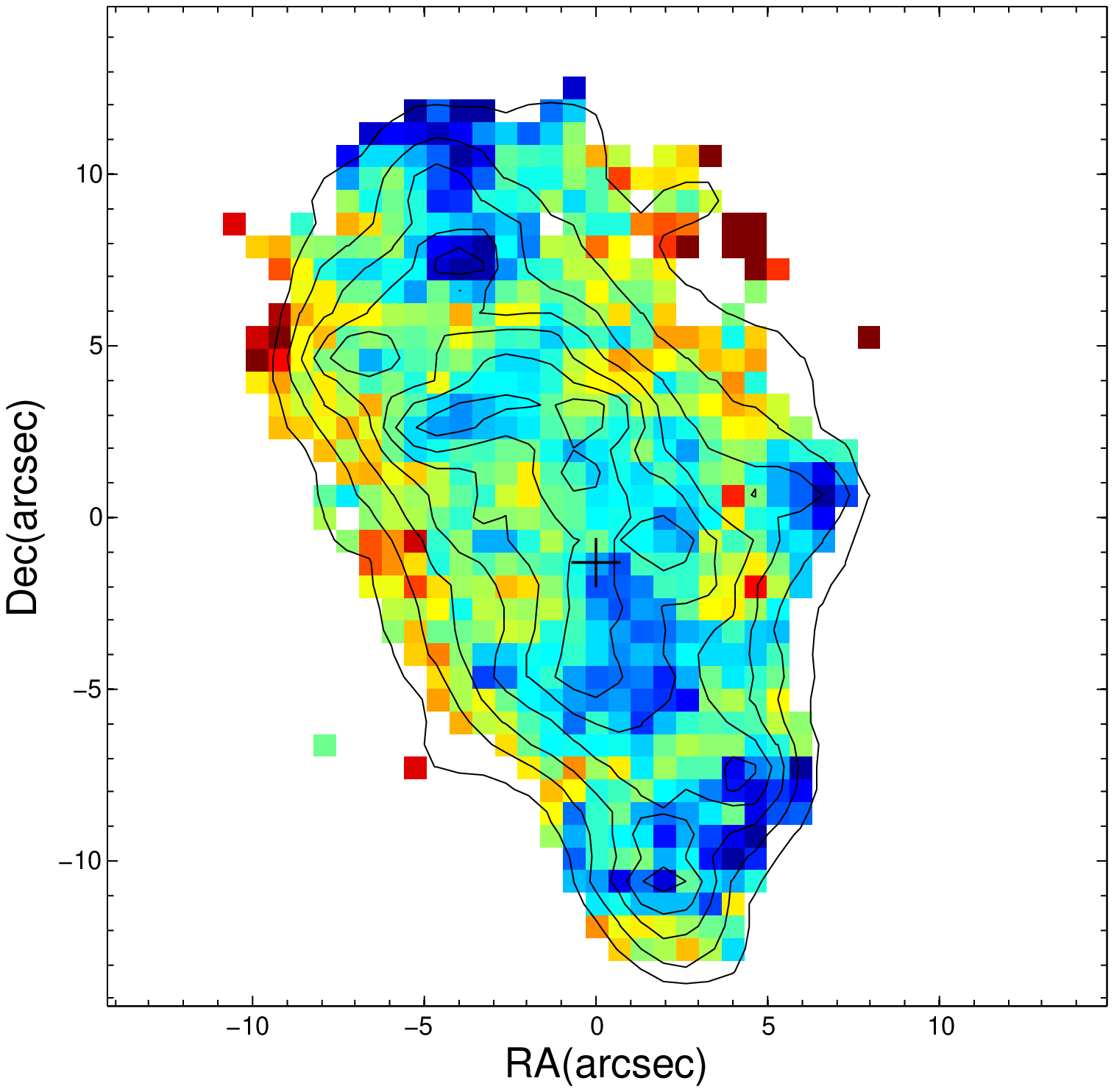}
\caption{Spatial localization of the spaxels in the diagnostic diagram [\ion{O}{iii}]~$\lambda5007$/\Hb{} versus
[\ion{S}{ii}]~$\lambda\lambda6717,\,6731$/\Ha{}. The color-code is the same 
as in Figure~\ref{Figure:diagnostic-spaxel}. The redder regions are 
those situated the above the maximum starburst line from \cite{Kewley2001}. }
\label{Figure:map-diagnostic} 
\end{figure}

Figure~\ref{Figure:map-diagnostic} shows the spatial location in the galaxy of the points plotted
in the [\ion{O}{iii}]~$\lambda5007$/\Hb{} versus [\ion{S}{ii}]~$\lambda\lambda6717,\,6731$/\Ha\
diagnostic diagram. Shocked regions are located  mainly in the galaxy outskirsts and in the
periphery of the SF regions (inter-knot areas), conforming with the idea of their being
generated by mechanical energy driven by stellar feedback. This is consistent  with  shocks
 caused  by the collective effect of massive stellar winds and SN remnants.  The same
behavior has been found for the other BCGs studied by means of IFU
\citep{Cairos2015,Cairos2017}.

\subsection{Integrated spectroscopy}
\label{Section:integrated}

We used the spatial information contained in emission line and continuum maps to delimitate
the nine main \ion{H}{ii}-regions seen in emission and the central continnum peak. For each
of these regions, we generated a high S/N spectrum by adding the spectra of their
corresponding spaxels, typically $\sim$10-20. To set the limits of the SF regions we worked
in the H$\alpha$ map, as this line, being the strongest recombination line in the optical,
shows the highest S/N and is, by comparison with H$\gamma$ and H$\beta$, not heavily
affected by stellar absorption \citep{Olofsson1995}. The nuclear region has been delimited
in the continuum frame. Because there is no definitive criterion  to set the limits of the
regions,  we adopted a pragmatic approach and integrated over a boundary that traces the
morphology of the cluster, keeping in mind that the minimun size is limited by the seeing.

\begin{figure}
\centering
\includegraphics[angle=0, width=1\linewidth]{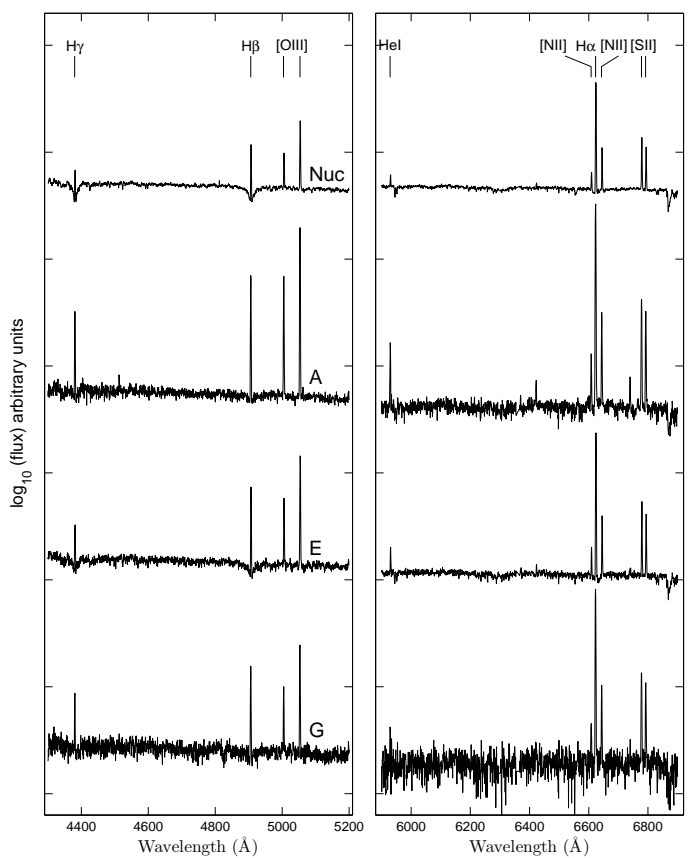}
\caption{Flux-calibrated spectra of three of the nine selected \ion{H}{ii} regions, and 
the nuclear region in Tololo~1937-423, 
in logaritmic units.}
\label{Figure:spectra} 
\end{figure}

The integrated spectra of the selected regions differ in their shapes and in the presence and/or strength
of several spectral lines. Knots~{\sc a} and {\sc b}, the two brighter  emitters in the central 
curvilinear feature (see Figure~\ref{Figure:Ha-contour}),  display both a typical nebular spectrum, with
prominent hydrogen recombination lines, [\ion{O}{iii}],  [\ion{N}{ii}] and [\ion{S}{ii}] in emission, 
in addition to a weak continuum, which slightly increases toward the blue wavelengths; small absorption wings are
visible around the Balmer lines in emission. To the south, the spectra of knots~{\sc c}, ~{\sc d,} and {\sc e}
show higher continua and clear absorption wings around H$\gamma$ and H$\beta$, revealing a major
contribution of the underlying stellar population.  Knots~{\sc f}, {\sc g}, ~{\sc h,} and {\sc i}, the
small SF regions detected at the filaments, display all prominent emission lines in addition to a weak and
featureless continuum. Finally, the spectrum of the nuclear region increases visibly toward blue
wavelengths,  H$\gamma$ and H$\beta$ display strong absorption wings, and absorption lines
characteristic of A stars such as CaI~$\lambda$4226 or MgII~$\lambda$4482 are clearly visible.  The high
ionization He~II~$\lambda$4686 \AA, typical feature of Wolf-Rayet (WR) stars, is not detected in any
knot. Figure~\ref{Figure:spectra} shows as an example the integrated spectrum of knots~{\sc a},
{\sc e}, {\sc g,} and the nuclear region in the wavelength range where spectral features have been
measured.

\subsubsection{Line fluxes and reddening correction}
\label{Section:fluxes}

Emission-line fluxes in each integrated spectrum were measured using the task {\sc splot}
within the {\sc iraf}\footnote{IRAF (Image Reduction and Analysis Facility) is a software system
for the reduction and analysis of astronomical data. It is distributed by the NOAO, which is
operated by the Association of Universities for Research in Astronomy, Inc., under cooperative
agreement with the National Science Foundation} environment. Fluxes were computed by fitting a
Gaussian plus a linear function to the line and continuum to all but the higher-order Balmer lines. 
The higher-order Balmer line fluxes in emission can be considerably affected by the underlying stellar absorption. To take
this effect into account, we fit the line profiles with two Gaussian functions in the cases in which the absorption wings are
clearly visible  -- H$\beta$ in all the spectra and H$\gamma$ in all but knots~{\sc g}, ~{\sc h,} and {\sc i}. In H$\alpha$,
where  the absence of visible absorption wings makes a reliable decomposition impossible,  we assumed the equivalent width in
absorption   to be equal to that in H$\beta$.  The equivalent width of H$\gamma$ and H$\beta$ in absorption in the cases where the
fit was made ranges between 2.0 and 5.3 \AA\  (see  Table~ \ref{tab:fluxes}),
which is in good agreement with the predictions of
evolutionary synthesis models \citep{Olofsson1995, GonzalezDelgado1999b}.

We derived the interstellar extinction coefficient from the H${\alpha}$/H${\beta}$ ratio
\citep{Osterbrock2006}. Although H$\gamma$ falls also in the observed spectral range, we did not
use this line  because its flux is considerably affected by the errors in the correction from underlying
absorption: the strength of the Balmer-lines series in emission decreases rapidly with decreasing
wavelength, whereas the equivalent width of the stellar absorption is roughly constant
\citep{Olofsson1995,GonzalezDelgado1999a}.  For the theoretical value of the H$\alpha$/H$\beta$ ratio,
we assumed   case~B approximation, with T=10000~K and N$_{e}\leq$100~cm$^{3}$.  The extinction law
from \cite{ODonnell1994} was adopted.

Reddening-corrected emission-line fluxes and interstellar extinction coefficients derived for each
integrated region are presented in Table~\ref{tab:fluxes}. Significant variations appear among the
interstellar  extinction coefficients for the different knots: the highest value  is found for the
nuclear region [C(H$\beta$)=0.76$\pm$0.02, E(B-V)=0.53$\pm$0.02], while knot~{\sc g} has  the
lowest coefficient  [C(H$\beta$)=0.18$\pm$0.05, E(B-V)=0.13$\pm$0.04]. The C(H$\beta$) value we
found for the nuclear region agrees well with the value derived from the integrated spectrum,
C(H$\beta$)=0.70, by \cite{Lagos2007}.


\subsubsection{Diagnostic line ratios, physical parameters, and abundances}

\label{Section:abundances}

The values of the diagnostic line ratios, and the derived  
electron densities (N$_{e}$), and the electron temperatures (T$_{e}$)
in the selected
regions of Tololo~1937-423 are shown in  Table~\ref{tab:diagnostic}. 



Electron densities were derived using the [\ion{S}{ii}]~$\lambda\lambda6717,\,6731$ line ratio 
\citep{Osterbrock2006}.  Electron temperatures were calculated from 
[\ion{O}{iii}]$\lambda$4363/($\lambda$4959+$\lambda$5007) in the knots where
[\ion{O}{iii}]$\lambda$4363 was reliably measured, that is, in  knots~{\sc a} and ~{\sc b}. Parameters
N$_{e}$ and T$_{e}$ were computed using the five-level atom {\sc fivel} program in the {\sc iraf
nebular} package \citep{deRobertis1987,ShawDufour1995}.  


The oxygen abundance could not be determined using the direct T$_{e}$-method, not even in the two
knots where [\ion{O}{iii}]$\lambda$4363 was measured,  because this method requires the measurement of
the [\ion{O}{ii}]~$\lambda$$\lambda$3727,3729 or, alternatively, the
[\ion{O}{ii}]~$\lambda$$\lambda$7320,7331 lines. [\ion{O}{ii}]~$\lambda$$\lambda$3727,3729 falls out
of the VIMOS spectral range, and the [\ion{O}{ii}]~$\lambda$$\lambda$7320,7331 lines,  although within
the observed range, fall very close to the edge of the spectrum, in  a zone that is highly affected by sky
residuals, which prevents obtaining reliable flux values. We therefore estimated the
oxygen abundances by adopting the new empirical method introduced by \cite{PilyuginGrebel2016}; these
authors provide a calibration particularly indicated for cases where 
[\ion{O}{ii}]~$\lambda$$\lambda$3727,3729 are not available, the so-called "S calibration": the
oxygen abundance is derived from the intensities of the strong lines
[\ion{O}{iii}]$\lambda$$\lambda$4957,5007, [\ion{N}{ii}]$\lambda$$\lambda$6548,6584, and
[\ion{S}{ii}]$\lambda$$\lambda$6717,6731.

The values derived for the different regions selected in Tololo~1937-423 are  also shown in Table~\ref{tab:diagnostic}. Their relative accuracy is 0.1~dex.  For comparison, abundances were also estimated using the [\ion{O}{iii}]/[\ion{N}{ii}]
indicator, as introduced in \cite{PettinPagel2004}, a calibration especially suited to be applied at higher redshifts.  All
regions present similar oxygen abundances, 12+log(O/H)$\sim$8.20 ($\approx$0.3~Z$\odot$\footnote{ solar abundances:
12+log(O/H)=8.69 \citep{Asplund2009}}). The values obtained by
applying the \cite{PettinPagel2004} calibration are systematically
higher than those obtained from \cite{PilyuginGrebel2016}, but are consistent within the uncertainties.

\begin{landscape}
\begin{table}
\footnotesize
\caption{Reddening-corrected line intensity ratios, normalized to H$\beta$, for the SF knots and the nuclear region in Tololo~1937-423.\label{tab:fluxes}}
\begin{tabular}{lcccccccccc}
\hline
Ion               &\multicolumn{1}{c}{A}&\multicolumn{1}{c}{B}      &\multicolumn{1}{c}{C}&\multicolumn{1}{c}{D}&\multicolumn{1}{c}{E}    &\multicolumn{1}{c}{F}&\multicolumn{1}{c}{G}& \multicolumn{1}{c}{H} & \multicolumn{1}{c}{I}&\multicolumn{1}{c}{Nuclear} \\
\hline                                                                                                                                                                                                                                                               
4340~H$\gamma$          &  0.487$\pm$0.017   &  0.503$\pm$0.016      &  0.502$\pm$0.016    &  0.510$\pm$0.024    &  0.411$\pm$0.022            &  0.520$\pm$0.036   &  0.424$\pm$0.031     & 0.543$\pm$0.086       & 0.355$\pm$0.036     &   0.438$\pm$0.022    \\
4363~[OIII]             &  0.031$\pm$0.006  &  0.029$\pm$0.008      &    ---              &  ---                &     ---                &   ---              &  ---               & ---                   &    ---              &   ---                   \\
4472~HeI               &  0.041$\pm$0.008   &  0.035$\pm$0.007      &    ---              &  ---                &     ---               &   ---               &  ---               & ---                   &      ---            &   ---                 \\
4861~H$\beta$           &  1.000            &  1.000               &  1.000               &  1.000               &  1.000                  &   1.000              & 1.000          &  1.000                  &  1.000                & 1.000 \\                                     
4959~[OIII]             &  0.976$\pm$0.019  &  0.955$\pm$0.018      &  0.715$\pm$0.015    &  0.502$\pm$0.017      &  0.670$\pm$0.022          &  0.804$\pm$0.030       &  0.542$\pm$0.034   &  0.732$\pm$0.036     & 0.625$\pm$0.040     &   0.585$\pm$0.018          \\
5007~[OIII]             &  2.865$\pm$0.039  &  2.872$\pm$0.046      &  2.056$\pm$0.035    &  1.518$\pm$0.035    &  1.980$\pm$0.048          &  2.346$\pm$0.069      &  1.570$\pm$0.072   & 2.011$\pm$0.071       & 1.625$\pm$0.073     &  1.754$\pm$0.036        \\
5875~HeI                &  0.111$\pm$0.006  &  0.107$\pm$0.006      &  0.094$\pm$0.006    &  0.086$\pm$0.009    &  0.105$\pm$0.012          &  ---                  &  ---              &  ---                 &  ---               &  ---  \\
6364~[OI]               &  0.024$\pm$0.004  &  0.030$\pm$0.005      &  0.041$\pm$0.005    &  0.040$\pm$0.008    &  0.032$\pm$0.018         &     ---       &       ---           &  ---                &     ---             &     ---                 \\
6548~[NII]              &  0.068$\pm$0.005  &  0.092$\pm$0.006      &  0.110$\pm$0.007    &  0.122$\pm$0.009    &  0.109$\pm$0.009          &  0.083$\pm$0.013       &  0.105$\pm$0.019   &   0.107$\pm$0.014    & 0.098$\pm$0.013     &    0.130$\pm$0.009     \\
6563~H$\alpha$           &  2.870$\pm$0.055  &  2.870$\pm$0.066      &  2.870$\pm$0.067    &  2.870$\pm$0.083   &  2.870$\pm$0.092           &  2.870$\pm$0.111     &  2.870$\pm$0.176   &   2.870$\pm$0.134      & 2.870$\pm$0.159     &          2.870$\pm$0.078     \\
6584~[NII]              &  0.215$\pm$0.007  &  0.281$\pm$0.010      &  0.350$\pm$0.011    &  0.356$\pm$0.015    &  0.334$\pm$0.015          &  0.281$\pm$0.017      &  0.318$\pm$0.028   &   0.273$\pm$0.020     & 0.338$\pm$0.026     &    0.374$\pm$0.013      \\
6678~HeI                &  0.033$\pm$0.005  &  0.031$\pm$0.005      &  0.026$\pm$0.005    &  0.021$\pm$0.008    &  ---                   &  ---                 &  ---               &   ---                 & ---                &      ---              \\
6717~[SII]              &  0.301$\pm$0.010  &  0.407$\pm$0.013      &  0.497$\pm$0.015    &  0.551$\pm$0.020    &  0.484$\pm$0.019          &  0.475$\pm$0.023      &  0.465$\pm$0.035   &    0.381$\pm$0.025     & 0.530$\pm$0.037    &   0.537$\pm$0.017      \\
6731~[SII]              &  0.218$\pm$0.008  &  0.307$\pm$0.011      &  0.366$\pm$0.010    &  0.391$\pm$0.016    &  0.348$\pm$0.017          &  0.343$\pm$0.020      &  0.340$\pm$0.029   & 0.294$\pm$0.020        & 0.362$\pm$0.026    &   0.387$\pm$0.015      \\
7065~HeI                &  0.017$\pm$0.004  &  0.020$\pm$0.005      &     ---             &      ---            &   ---                  &     ---             &  ---              &    ---                 & ---                &       ---                         \\
7136~[ArIII]            &  0.070$\pm$0.005  &  0.069$\pm$0.005      &  0.064$\pm$0.004    &  ---                &  0.062$\pm$0.009          &  ---                   &  ---            &  ---                   & ---                 &    ---   \\
\hline
F$_{H\beta}$            &  56.52$\pm$3.23     &  50.6$\pm$2.78        &  104.2$\pm$9.4     &  87.63$\pm$15      &  48.9$\pm$5.6            &  15.85$\pm$1.4     &  9.4$\pm$1.0          & 25.91$\pm$6.6        &  15.13$\pm$2.65  &  179.66$\pm$37.8      \\
C$_{H\beta}$            &  0.403$\pm$0.016  &  0.313$\pm$0.019      &  0.504$\pm$0.019    &  0.667$\pm$0.024   &  0.470$\pm$0.027           &  0.288$\pm$0.032    &  0.182$\pm$0.051   & 0.625$\pm$0.039          &  0.422$\pm$0.046 &    0.760$\pm$0.022           \\
W(H$\gamma$)$_{ab}$     &  2.1                &  3.8                &          4.5       &     4.0             & 3.1                     &    2.4           &      ---           & ---                  & ---              &         4.8                       \\
W(H$\beta$)$_{ab}$      &  2.2              &  2.0                  &      4.6           &     3.7               & 3.3                         &   2.1              &       2.2               & 2.9           & 2.9                  &        5.3                        \\
$A_{V}$                 &  0.870$\pm$0.034  &  0.676$\pm$0.041      &   1.088$\pm$0.042   &  1.442$\pm$0.052   &  1.015$\pm$0.058           &  0.622$\pm$0.070     & 0.393$\pm$0.110    & 1.350$\pm$0.084         & 0.912$\pm$0.099  &    1.642$\pm$0.005     \\                                                                                                                                                                                            
E(B-V)                  &  0.281$\pm$0.011 &  0.218$\pm$0.013      &   0.351$\pm$0.013  &  0.465$\pm$0.017    &  0.328$\pm$0.019         &  0.201$\pm$0.022       &  0.127$\pm$0.036   & 0.435$\pm$0.027         & 0.294$\pm$0.032  &      0.530$\pm$0.016      \\
\hline
\end{tabular}
\end{table}
Notes.- Reddening-corrected line fluxes normalized to F(H$\beta$)=1.
The reddening-corrected H$\beta$
flux (in units of 10$^{-16}$erg~s$^{-1}$~cm$^{-2}$), 
the
interstellar extinction coefficient,C$_{H\beta}$, 
and values of the equivalent width in absorption for H$\gamma$ and H$\beta$ 
are also provided in the table. A$_{V}$=2.16$\times$C(H$\beta$) and
$E(B-V)$=0.697$\times$C(H$\beta$) \citep{Dopita2003}.
\end{landscape}

\begin{landscape}
\begin{table}
\footnotesize
\caption{Diagnostic line ratios, physical parameters and abundances for the SF and the nuclear region in Tololo~1937-423.\label{tab:diagnostic}}
\begin{tabular}{lccccccccccc}
\hline
Parameter                                               &\multicolumn{1}{c}{A}       &\multicolumn{1}{c}{B}      &\multicolumn{1}{c}{C}        &\multicolumn{1}{c}{D}     &\multicolumn{1}{c}{E}       &\multicolumn{1}{c}{F}         &\multicolumn{1}{c}{G}        & \multicolumn{1}{c}{H}    &\multicolumn{1}{c}{I}            &\multicolumn{1}{c}{Nuclear} \\
\hline
$[\ion{O}{iii}]~\lambda5007/\Hb$                         & 2.865$\pm$0.039 & 2.872$\pm$0.046 & 2.056$\pm$0.035 & 1.518$\pm$0.035 & 1.980$\pm$0.048 & 2.346$\pm$0.069 & 1.570$\pm$0.072 & 2.012$\pm$0.071 & 1.625$\pm$0.073 & 1.754$\pm$0.036 \\
$[\ion{N}{ii}]~\lambda6584/\Ha$                          & 0.075$\pm$0.002 & 0.098$\pm$0.003 & 0.122$\pm$0.003 & 0.124$\pm$0.004 & 0.116$\pm$0.004 & 0.098$\pm$0.005 & 0.111$\pm$0.007 & 0.095$\pm$0.006 & 0.118$\pm$0.006 & 0.130$\pm$0.003 \\
$[\ion{S}{ii}]~\lambda\lambda6717\,6731/\Ha$             & 0.181$\pm$0.003 & 0.249$\pm$0.004 & 0.301$\pm$0.004 & 0.328$\pm$0.006 & 0.290$\pm$0.006 & 0.285$\pm$0.007 & 0.280$\pm$0.010 & 0.235$\pm$0.008 & 0.311$\pm$0.009 & 0.322$\pm$0.005 \\
N$_{e}$   (cm$^{-3}$)                                     &   $<$100                  &   $<$100                  &$<$100                   &  $<$100                &     $<$100               &  $<$100                  &    $<$100                    &    125                    & $<$100                     &      $<$100          &   \\
T$_{e}$  (K)                                             &    11950                  &  11700                 &    ---                     &      ---                 &       ---                &         ---              &   ---                    &     ---                      &     ---                       &   ---                      \\
12+log(O/H)$^{1}$                                        &    8.17                   &  8.19                     &  8.21                      &   8.20                 & 8.21                      &  8.15                       &  8.20                    &     8.20                       & 8.19                   &   8.23         \\
12+log(O/H)$^{2}$                                        &    8.22                   &  8.26                     &  8.34                      &   8.38                 & 8.34                      &  8.29                       &  8.36                &    8.30                       &  8.36                 &   8.37          \\
\hline
\end{tabular}
Notes.-  (1) Abundances estimated from \cite{PilyuginGrebel2016} ---uncertainties are within
0.1~dex; (2) Abundances derived following \cite{PettinPagel2004} ---uncertainties are within
0.25~dex.  
\end{table}
\end{landscape}

\normalsize

\subsection{Ionized gas kinematics}

\begin{figure*}
\centering
\includegraphics[angle=0, width=0.8\linewidth]{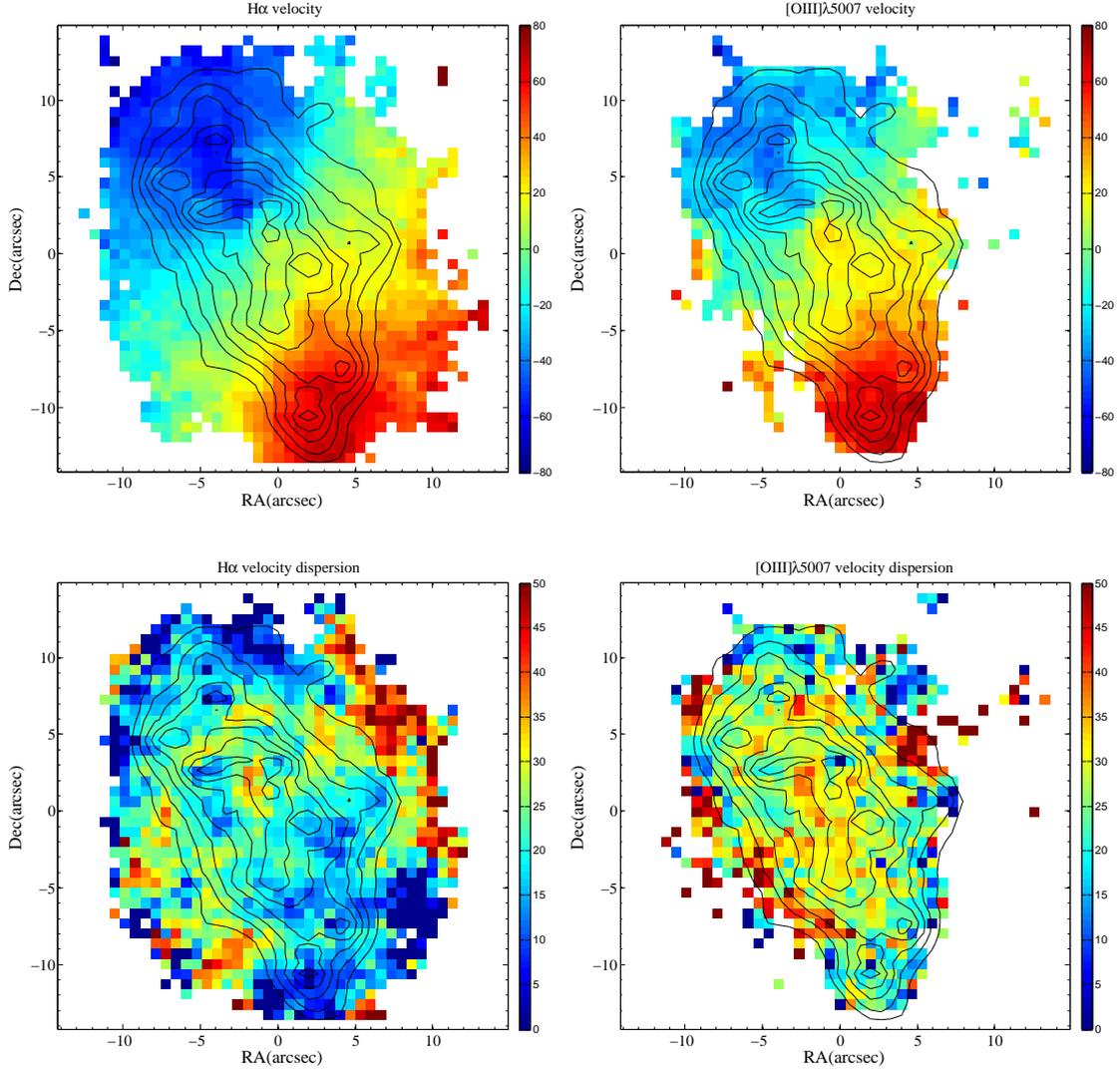}
\caption{H$\alpha$ and [\ion{O}{iii}]$\lambda$5007 line-of-sight velocity
fields (at the top) and velocity dispersion maps (at the bottom).}
\label{Figure:velocitymaps} 
\end{figure*}

We investigated the kinematics of the emitting gas in Tololo~1937-423 from the two brightest emission lines, namely, H$\alpha$
and [\ion{O}{iii}]$\lambda$5007. Figure~\ref{Figure:velocitymaps} (top panels) presents the  H$\alpha$ and
[\ion{O}{iii}]$\lambda$5007 line-of-sight (LOS) velocity maps measured from the Doppler shift of the line profile centroid
relative to the galaxy systemic velocity. The H$\alpha$ velocity field is more extended than that of the [\ion{O}{iii}]$\lambda$5007
line because of the better H$\alpha$ S/N, but both maps exhibit the same large-scale pattern: an overall rotation along an axis
oriented southeast-northwest, with the regions situated northeast moving toward us and  those placed southwest moving away. 
The amplitude of the H$\alpha$ velocity field is  $\sim$125~km~s$^{-1}$.

The LOS velocity dispersion was derived from the width of the Gaussian fit to the line profile after
accounting for the instrumental broadening.  The H$\alpha$ and [\ion{O}{iii}]$\lambda$5007 velocity
dispersion maps both show a clear spatial pattern (Figure~\ref{Figure:velocitymaps}, bottom panels):
relatively low velocity dispersions (10~km~s$^{-1}$ $\leq$ $\sigma$ $\leq$ 15~km~s$^{-1}$) are found in
the SF regions,  whereas the values increase up to 45-50~km~s$^{-1}$ at the northwest and southeast
galaxy edges and in the space between the SF knots. Similar patterns have been found in the few other
blue compact and dwarf galaxies for which 2D maps of the ionized gas velocity dispersion are available
\citep{Bordalo2009,Moiseev2012, Moiseev2015,Cairos2015,Cairos2017,Lagos2016}.
\cite{Moiseev2015} argued that the ionized gas turbulent motions in dwarfs do not reflect virial
motions, but are instead connected with stellar feedback, that
is, with the energy injected into the ISM by
stellar winds and SN explosions. Numerical simulations  yield similar conclusions \citep{Dib2006}.

\smallskip

Applying the commonly used tilted-ring analysis  \citep{Rogstad1974,Begeman1987,Begeman1989},  we
can derive the galaxy rotation curve from the observed velocity field.  This method assumes that the galaxy velocity field is only due to rotation and that the disk is made of several
concentric (not necessarily coplanar) rings, each of them characterized by its rotational
velocity, $v_{rot}$, radius, $r$, and inclination, $i$.  The LOS velocity field in the plane of
the sky is then\begin{equation} \label{eqTilted1} v(x,y) = v_{sys} + v_{rot} \sin i \cos \theta,
\end{equation} \noindent where $v_{sys}$ is the systemic velocity and $\theta$ the azimuthal angle
from the major axis  in the plane of the galaxy. The azimuthal angle is related to the center of
the ring ($x_0,y_0$), the position angle in the sky ($PA$) and $i$ as follows: \begin{equation}
\label{eqTilted2} \cos\theta = \frac{-(x-x_0)\sin(PA) + (y-y_0)\cos(PA)}{r}, \end{equation}
\begin{equation} \label{eqTilted3} \sin\theta = \frac{-(x-x_0)\cos(PA) - (y-y_0)\sin(PA)}{r\sin
i}. \end{equation}

The parameters $v_{rot}$, $i$, $PA$, as well as the position of the center $(x_0, y_0)$ and
$v_{sys}$ are computed through a nonlinear least-squares fit of the observed velocity field to
the model (Equation~\ref{eqTilted1}).

While $v_{rot}$, $i$, and $PA$ take different independent values for each ring,  $(x_0, y_0)$ and
$v_{sys}$ are common. Generally, these two values are determined iteratively (e.g., Begeman 1989;
Elson et al. 2010): all parameters are fit independently for each ring, and after this, the
common $(x_0, y_0)$ and $v_{sys}$ are obtained by averaging over the values of the individual rings.
Next, the fit is repeated, but now with $(x_0, y_0)$ and $v_{sys}$ fixed.  However,
proceeding in this way, we have detected convergence problems, which are most probably due to some degeneracy between
$(x_0, y_0)$ and $v_{sys}$, so we employed an alternative method. 
We found it much more robust to first determine  $(x_0, y_0)$ and
$v_{sys}$ independently by applying the following simple symmetry argument: the tilted-ring velocity
field [Equations~\ref{eqTilted1}-\ref{eqTilted3}] is (anti)~point-symmetric, that is, any two points
$P$ and $Q$ at the same distance from the center $(x_0, y_0)$ but opposite position angles, $PA$
and $PA+180^\circ$, have velocities $v_{P,Q}=v_{sys}\pm \Delta$ and hence, the mean LOS velocity
of the pair is $v_{sys}$. We computed then $(v_P+v_Q)/2$ across the galaxy and  defined the galaxy
center as the position in which the dispersion of the means $(v_P+v_Q)/2$ is minimal, that is,  we
created an algorithm that finds $(x_0, y_0)$ and $v_{sys}$ by minimizing  \begin{equation}
\frac{1}{N}\sum_P \left(\frac{v_P + v_Q}{2} - v_{sys}\right)^2 \end{equation} \noindent using a
least-squares algorithm for all $N$ points $P$ in the galaxy and their point-antisymmetric
Q---with respect to $(x_0, y_0)$---. The algorithm starts at an initial guess and converges to the
center of the velocity field.

\begin{figure}
\centering
\includegraphics[angle=0, width=0.9\linewidth]{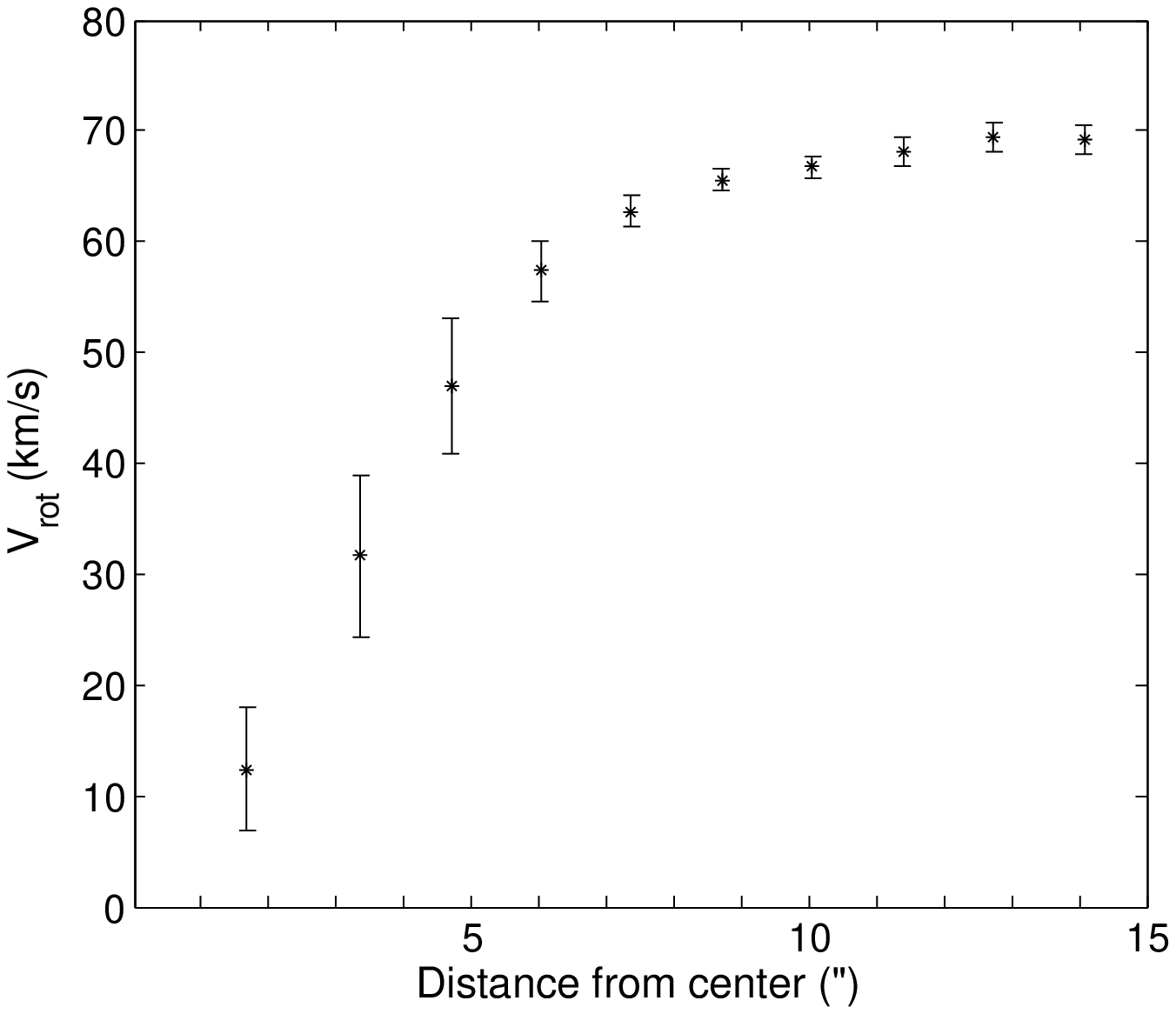}
\caption{Measured rotation curve of the ionized gas in Tololo1937-423 from H$\alpha$. }
\label{Figure:rotcur} 
\end{figure}

The rotation curve of Tololo~1937-423,  inferred from the $H_\alpha$ velocity field, is shown in
Figure~\ref{Figure:rotcur}. The curve increases rapidly in the central areas of the galaxy, and flattens at
distances $\geq 10"$ ($\sim 2$~kpc) to $v_{rot}=70\pm7$~km~s$^{-1}$.
This is calculated from the average of the last
three data points. From $v_{rot}$ we can estimate the dynamical mass enclosed in a given radius, r, using
\begin{equation}
\label{eqmassdy}
M(r) = \frac{v_{rot}^{2}(r)~r}{G},
\end{equation}
\noindent where G is the gravitational constant. Equation~\ref{eqmassdy} assumes a spherical mass
distribution, that is, a dark-matter-dominated galaxy. This is
a reasonable assumption that takes into account
the findings for most studied dwarfs (see, e.g., \citealp{Carignan1989}). From here we derived a  
dynamical
mass of M=2.9$\times$10$^{9}$~M$_{\odot}$ for Tololo~1937-423.

\section{Discussion} 

\label{discuss}

\begin{table*}
\caption{H$\alpha$ derived quantities for the individual SF regions identified in Tololo1937-423.}
\begin{center}
\begin{tabular}{|c|c|c|c|c|c|c|c|}
\hline
Knot    &  F(H$\alpha$)                    & log[L(H$\alpha$)]            & W(H$\alpha$)           &  SFR                                 & N$_{LyC}$                        &   Age \\
        & (10$^{-16}$~erg cm$^{-2}$s$^{-1}$) & (erg~s$^{-1}$)     &         (\AA)           & 10$^{-2}$M$\odot$ ~yr$^{-1}$   & (10$^{51}$~photons~s$^{-1}$)          &   (Myrs)\\\hline\hline
A       &  162$\pm$4                        &      39.52          &  232            &  2.60                     &  2.41                          &             5.4\\
B       &   145$\pm$4                       &      39.47          &  158            & 2.33                      &  2.15                          &              5.9  \\     
C       &  299$\pm$11                       &      39.78          &73                       & 4.79                             &  4.43                        &             6.6   \\
D       &  251$\pm$16                       &      39.71          &54                       & 4.03                             & 3.73                 & 7.0   \\
E       &  140$\pm$7                        &      39.45         &51                        & 2.25                             & 2.08                 &  7.1  \\
F       &  45$\pm$2                         &      38.96          &351              & 0.73                             & 0.67                    & 5.0 \\
G       &  27$\pm$2                         &      38.74          & 97              & 0.43                             & 0.40                   & 6.3 \\
H       &  74$\pm$7                         &      39.18         & 169              & 1.19                             & 1.1                    & 5.8\\
I       &  43$\pm$4                         &      38.94         &162                       & 0.70                             & 0.64                 & 5.9 \\\hline
\end{tabular}       
\end{center}
Notes: H$\alpha$ fluxes were corrected for interstellar extinction using the values provided in Table~\ref{tab:fluxes}
\label{SF-KnotI}
\end{table*}

Tololo~1937-423 is currently undergoing an extended SF episode, as revealed by its emission line
maps. The ionized gas traces the current SF because only short-lived OB stars are able to produce
photons energetic enough to ionize hydrogen. In the H$\alpha$ emission-line maps we have identified
nine main SF knots, aligned northeast-southwest, and stretching to galactocentric distances of up to
2.5~kpc (see Section~\ref{Section:integrated}). Probing their spatial and temporal distribution,
that is, examining their morphological pattern and estimating their ages, will give us information
on the mechanism(s) that trigger and propagate the recent SF.

Hydrogen recombination lines are used to constrain the properties of \ion{H}{ii}-regions, as their
flux is proportional to the number of ionizing photons radiated by the central stars. In particular,
H$\alpha$ is an excellent diagnostic line in the nearby Universe: its luminosity is a robust SFR
calibrator \citep{Kennicutt1998} and its equivalent width is very highly sensitive to the starburst
age \citep{Leitherer1999}. 

Table~\ref{SF-KnotI} compiles the H$\alpha$ fluxes, luminosities, and equivalent widths for selected
SF regions in Tololo~1937-423. All knots have  H$\alpha$ luminosities typical of supergiant
\ion{H}{ii}-regions, but knots close to the galaxy center are brighter than knots located in the
filaments. We derived the SFR and the number of ionizing photons (N$_{LyC}$) from the H$\alpha$
luminosity following \cite{Osterbrock2006} (Columns~5 and 6 in Table~\ref{SF-KnotI}).

\begin{figure}
\centering
\includegraphics[angle=0, width=0.8\linewidth]{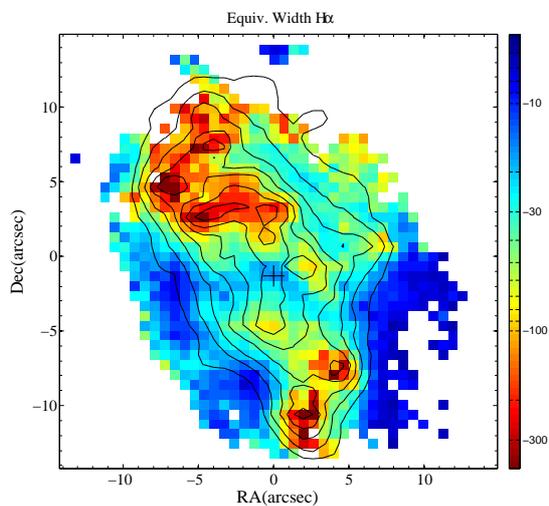}
\caption{H$\alpha$ equivalent width map for Tololo~1937-423 with the H$\alpha$ contours
overplotted.}
\label{Figure:eqwha} 
\end{figure}

We estimated the ages of the SF knots by comparing their H$\alpha$ equivalent widths with the
predictions of the {\sc Starburst~99} evolutionary synthesis models \citep{Leitherer1999}.
Adopting a metallicity z=0.008 (the value closest to the metallicity derived from the emission-line fluxes), we found that the knot equivalent widths can be reproduced with an
instantaneous burst of SF, a Salpeter initial mass function (IMF) with an upper mass limit
of 100 M$\odot$,  and  ages of between  5.0 and 7.1~Myrs (last column 
in Table~\ref{SF-KnotI}).

It is known, however, that the measured Balmer-line equivalent widths can decrease as a result of A-F
stars absorption and/or a high continuum from an older stellar population
\citep{Levesque2013}. The H$\alpha$ measurements in  Table~\ref{SF-KnotI} are corrected  for
stellar absorption (see Section~\ref{Section:fluxes}), but not for an older
star continuum. A correction for a higher continuum would require analyzing the spectra
in terms of population synthesis, or alternatively, the structural modeling of the underlying
population using additional data (e.g., deep broadband imaging), and this is beyond the scope of
this work. Thus, the derived ages must be understood as an upper limit. 
\cite{Cairos2002,Cairos2007} modeled the LSB component of the BCDs Mrk~370 and Mkr~35 and
estimated that the falling off in the H$\alpha$ equivalent width associated with a higher
continuum  could be as high as a 70\% and that this was additionally largely
dependent on the spatial position. The uncertainties in the derived ages due to this
variations are    
$\sim$1-1.5~Myr.

Since the H$\alpha$ equivalent width decreases almost monotonically with starburst age,
the H$\alpha$ equivalent width depicts the age of the SF regions (Figure~\ref{Figure:eqwha}). 
Younger knots (i.e., larger equivalent widths) are found at larger galactocentric distances,
while older knots (i.e., smaller equivalent widths)  lie closer to the continuum peak. 
This indicates that dilution due to a high continuum is the  apparent origin of the gradient in
equivalent widths, and consequently, of the derived ages, and it supports the hypothesis that the
\ion{H}{ii} regions in Tololo~1937-423 depict a simultaneous SF episode, most likely triggered
by a large-scale mechanism that was active in the galaxy about 5-7~Myr ago. The nearly constant 
oxygen abundance found for all the clumps also indicates that all of them  belong to
the same SF event.

The point symmetric distribution of ionized gas around the continuum peak suggests  
triggered SF.  Large holes and shells are frequently found in the \ion{H}{i} and ionized gas
distribution of dwarf galaxies  \citep{Puche1992,Staveley1997,Walter1999,Kim1999}. Holes in
\ion{H}{i} are thought to be created by feedback from massive stars: the supersonic flows
generated by distant SN blast waves and/or stellar winds sweep up a shell of shocked
neutral gas 
\citep{Weaver1977,McCray1987,TenorioTagle1988,Weisz2009a,Weisz2009b,Warren2011,Bagetakos2011}. 
Evidence is accumulating that  in the rims of these bubbles, new (secondary) star formation can
occur \citep{Lozinskaya2002,Cannon2005,Egorov2014,Egorov2017,Cairos2017}. In line with these
findings,  the ionized gas emission in Tololo~1937-423 could be tracing a secondary SF episode.

To validate the likelihood of a scenario of propagated SF,  we have to  find the remmant
of the OB association whose massive stars caused the  blast wave. Then,  we must proof
that both SF events, the current and the previous
SF episode, could indeed be causally
related.

The obvious progenitor candidate in Tololo~1937-423 is the stellar cluster (or clusters) that
generates the continuum peak. Broadband surface photometry shows an apparent color gradient in
the galaxy \citep{Doublier1999,GildePaz2005}. \cite{GildePaz2005} derived a color
$(B-R)$=1.21$\pm$0.11 for the LSB underlying host, indicative of ages of several Gyr
\citep{Vazdekis1996}, whereas from their color profile we infer $(B-R)\sim$0.8 at the galaxy
center. We showed in Section~\ref{Section:extinction} that the interstellar extinction
also peaks in this region, which means that the actual $(B-R)$ is indeed even bluer. This 
color gradient is most probably evidence of an age gradient.  Moreover, the strong  absorption
in H$\beta$ and H$\gamma$ in the nuclear spectrum (see Section~\ref{Section:integrated} and
Figure~\ref{Figure:spectra}) reveals the presence of A-F intermediate-age stars. The value 
of the equivalent width in absorption of higher-order Balmer lines has been shown to
effectively  constrain the age of a poststarburst
\citep{Olofsson1995,GonzalezDelgado1999a,GonzalezDelgado1999b}. By comparing  H$\gamma$ and
H$\beta$ equivalent widths in absorption with the predictions of the  synthesis models by
\cite{GonzalezDelgado1999b}, we estimated the ages for the nuclear region of $\sim$13 and
20~Myr, assuming  an instantaneous burst (IB) and solar metallicity (Z$\odot$) and
0.05~Z$\odot$, respectively. Since we are integrating over a region of diameter
$\sim$600~pc  that probably contains several stellar clusters, the IB aproximation is
likely  too simplistic and the derived ages are lower limits.   Using the
model predictions for a continuous SF burst, we obtained older ages of $\sim$40-80~Myr.

We found that the stellar population in the central region of the galaxy is  several tens of
Myr old (13-80~Myr), while the more extended and recent SF episode is  $\sim$5-7~Myr old.
These ages, together with the distance between the different SF episodes,  imply velocities
of 20 to 100~km~s$^{-1}$ for the blast wave caused by the SN explosions from
the central post-starburst, to  be able to ignite the current SF,  in good agreement with the
expansion velocities found for shells in dwarf galaxies
\citep{Walter1999,Egorov2014,Egorov2017}.  Hence, the morphology and age distribution of
the SF episode  both indicate triggered SF in Tololo~1937-423, with the most recent SF episode
triggered by the collective effects of stellar winds and SN explosions from the central
post-starburst.


Other observables also favor triggered SF in Tololo~1937-423. The ionized gas morphology,
with filaments and curvilinear structures, is characteristic of an ISM perturbed by the action
of stellar winds and SN explosions. The velocity dispersion pattern,  with  turbulent motions
increasing toward the periphery of the SF regions and at the  galaxy edges, can be naturally
explained if these high dispersions are due to the mechanical energy injected into the gas by
stellar feedback. The presence of shocked regions in the galaxy also conforms to triggered SF,
since the most viable mechanism to drive shocks in star-forming galaxies are SN explosions and
massive stellar winds \citep{Shull1979,Allen1999}.


Recent Hubble Space Telescope analyses of resolved stellar populations on
nearby dwarf galaxies have shown that the starburst episodes in these galaxies are indeed much
longer than previously thought, lasting up to $\sim$200-400~Myr
\citep{McQuinn2009,McQuinn2010a,McQuinn2010b}. Feedback from massive stars 
in dwarfs and, more specifically, new SF triggered by the cumulative effects of stellar winds
and SN explosions, would naturally explain such temporally extended starbursts.


\section{Summary}

We presented results from a broad analysis of the BCG Tololo~1937-423 based on integral field
observations. We mapped the central 5.4$\times$5.4~kpc$^{2}$ of the galaxy in the 4150--7400\,\AA\
spectral range with a spatial resolution of $\sim$133~pc using VIMOS at the VLT, and probed  its
morphology, stellar content, nebular ionization and excitation properties, as well as the
kinematics of its ionized gas. Our main conclusions are the following.

$\bullet$ The SF in Tololo~1937-423 is spatially extended. The current SF episode is taking place
in various knots, which reach galactocentric distances of up to 2.5~kpc. The brighter SF clumps
are close to the center of the galaxy, while fainter clumps are located in filaments stretching
northeast and southwest. We delimited nine main SF regions in the galaxy and produced their
integrated spectrum, from which  reliable physical parameters and oxygen abundances were
computed.   We found that all clumps  present similar abundances (12+log(O/H)=8.20$\pm$0.1); we did
not find significant variations on scales of kpc.

$\bullet$ Tololo~1937-423 displays distinct patterns in their emission line and continuum maps.
While emission line maps are extended and knotty, the continuum map has a single peak, situated
roughly at the galaxy center, which does not spatially coincide with any clump in emission lines.
This implies at least two relatively recent SF episodes. Using  evolutionary
synthesis models, we estimate ages of $\sim$5-7~Myr for the ongoing starburst, whereas the stellar
population in the central region of the galaxy is found to be tens of Myr old (13-80~Myr).

$\bullet$ The galaxy shows an inhomogeneous dust distribution, with the extinction peak reached
roughly at the position of the continuum peak. Here, the color excess is E(B-V)$\sim$0.71.

$\bullet$  Although photoionization by hot stars is the dominant excitation and ionization
mechanism in the galaxy, shocks are playing a major role in the galaxy outer regions and in the
periphery of the SF regions (inter-knot areas). 

$\bullet$  The LOS ionized gas velocity field of Tololo~1937-423 displays ordered motions, with an
overall rotation around an axis oriented SE-NW. The amplitude of the H$\alpha$ velocity field is
$\sim$125~km~s$^{-1}$. The rotation curve, inferred from the H$\alpha$ velocity field, increases
rapidly in the central areas of the galaxy and flattens at distances $\geq 10"$ ($\sim2$~kpc) to
a value $V_{rot}=70\pm7$~km~s$^{-1}$.  Assuming that the galaxy is dark matter dominated, we
derived a dynamical mass M=2.9$\times$10$^{9}~M_{\odot}$.

$\bullet$ The LOS velocity dispersion maps of Tololo~1937-423  show 
relatively low velocity dispersions (10~km~s$^{-1}$ $\leq$ $\sigma$ $\leq$ 15~km~s$^{-1}$)  in the SF
regions,  and values that increase up to 45-50~km~s$^{-1}$ at the northwest and southeast
galaxy edges and in the space between the SF knots.

$\bullet$ The morphology of the galaxy and the two distinct SF
episodes suggest a scenario
of triggered SF. Ages derived for the different SF episodes, $\sim$13-80~Myr for the central
post-starburst and 5-7~Myr for the ongoing SF,  are  consistent with star-induced SF, with the most
recent SF episode caused by the collective effect of stellar winds and SN explosions from 
the central
post-starburst. The velocity dispersion pattern (higher velocity dispersions at the edges of the SF
regions) and the shocked regions in the galaxy also support this scenario.

\begin{acknowledgements} LMC acknowledges support from the Deutsche Forschungsgemeinschaft
(CA~1243/1-1). The authors are very grateful to N. Caon and P. Weilbacher, members of the IFU-BCG project.
We also thank R. Manso-Sainz and Oliver Mettin for many constructive discussions. 
Further thanks go to R. Mettin for a careful reading
of the manuscript. 
This research has made use of the NASA/IPAC Extragalactic
Database (NED), which is operated by the Jet Propulsion Laboratory, Caltech,
under contract with the National Aeronautics and Space Administration. 
\end{acknowledgements}

\bibliographystyle{aa}
\bibliography{vimos}

\end{document}